\documentclass{article}
\usepackage[utf8]{inputenc}
\usepackage{color}
\usepackage{ragged2e}
\usepackage{graphicx}
\graphicspath{{figures/}}
\usepackage{booktabs}
\usepackage{multicol}
\usepackage{multirow}
\usepackage[section]{placeins}
\usepackage[table]{xcolor}
\usepackage{makecell}
\usepackage{pdflscape}
\usepackage[margin=1.5in]{geometry}
\usepackage[T1]{fontenc}

\usepackage{amsmath,amssymb}
\usepackage{caption}
\usepackage{natbib}
\setcitestyle{authoryear,open={(},close={)}}
\usepackage{authblk}
\usepackage{silence}
\title{Enhancing Poverty Targeting with Spatial Machine Learning: An application to Indonesia}

\author[1]{Rolando Gonzales Martinez}
\author[2]{Mariza Cooray}
\affil[1]{University of Groningen. Contact: r.m.gonzales.martinez@rug.nl}

\date{January 2025}

\begin{document}

\maketitle

\begin{abstract}
\noindent This study leverages spatial machine learning (SML) to enhance the accuracy of Proxy Means Testing (PMT) for poverty targeting in Indonesia. Conventional PMT methodologies are prone to exclusion and inclusion errors due to their inability to account for spatial dependencies and regional heterogeneity. By integrating spatial contiguity matrices, SML models mitigate these limitations, facilitating a more precise identification and comparison of geographical poverty clusters. Utilizing household survey data from the Social Welfare Integrated Data Survey (DTKS) for the periods 2016–2020 and 2016–2021, this study examines spatial patterns in income distribution and delineates poverty clusters at both provincial and district levels. Empirical findings indicate that the proposed SML approach reduces exclusion errors from 28\% to 20\% compared to standard machine learning models, underscoring the critical role of spatial analysis in refining machine learning-based poverty targeting. These results highlight the potential of SML to inform the design of more equitable and effective social protection policies, particularly in geographically diverse contexts. Future research can explore the applicability of spatiotemporal models and assess the generalizability of SML approaches across varying socio-economic settings.
\end{abstract}

\newline
\hspace*{0.55cm}\textit{Keywords:} Proxy means testing, spatial machine learning, poverty targeting 

\section{Introduction}

Targeting methods are essential for the effective implementation of social safety nets in developing countries. However, no singular method is universally applicable to poverty targeting. \citet{coady2004targeting} conducted an extensive review of various targeting methods, emphasizing the need for both efficiency and equity in beneficiary selection. Their analysis revealed that different methods exhibit distinct levels of targeting accuracy, often presenting trade-offs between \textit{undercoverage} (exclusion of eligible beneficiaries) and \textit{leakage} (inclusion of ineligible beneficiaries).

Proxy Means Testing (PMT) is widely used to identify beneficiaries of social assistance programs. PMT relies on observable household characteristics to estimate welfare levels, targeting households with low average per capita consumption. However, PMT models often suffer from issues related to data accuracy and inherent biases \citep{gazeaud2020proxy}. Econometric approaches such as simple mean regressions may unintentionally exclude a substantial portion of impoverished households, thereby reducing the impact of social welfare initiatives \citep{brown2018poor}.

Advancements in machine learning (ML) offer promising opportunities to improve targeting accuracy and are increasingly regarded as the future of poverty targeting \citep{heidelberg2024pmt}. For instance, \citet{aiken2022machine} shows that ML models, particularly those using mobile phone data, can significantly enhance the precision of humanitarian aid allocation, resulting in more efficient resource distribution. Similarly, \citet{mcbride2018retooling} shows that incorporating out-of-sample validation techniques with ML substantially improves the accuracy of poverty targeting tools.

Despite these technological advancements, variability in targeting efficiency persists across countries and regions. \citet{alatas2012targeting} reveals notable disparities in targeting effectiveness across diverse demographic and socioeconomic contexts. These findings underscore the need for geographically tailored targeting methodologies, including those employing ML techniques.

Given the critical role of regional variability in poverty targeting and the emerging application of machine learning, this study proposes including \textit{spatial machine learning} (SML) to account for geographical factors in PMT models. Using Indonesia's \textit{Data Terpadu Kesejahteran Sosial} (DTKS) survey - the primary registry for identifying households eligible for social transfers by the Indonesian Ministry of Social Affairs, we develop and evaluate SML-PMT models. We utilize both supervised and unsupervised SML models to minimize \textit{exclusion errors} (undercoverage, misclassification of poor households as non-poor) and \textit{inclusion errors} (leakage, misclassification of non-poor households as poor).

Indonesia’s unique geographic composition as an archipelago of thousands of islands necessitates the incorporation of spatial location in PMT algorithms based on SML. \citet{nikparvar2021machine} argue that the properties of spatial data are often overlooked or inadequately addressed in ML models. Ignoring spatial autocorrelation within Indonesia's regions can lead to ML-PMT models that perform adequately on training data but fail to generalize predictions across spatially correlated areas \citep{meyer2019importance}. Our findings indicate that explicitly incorporating spatial properties at the sub-national level significantly enhances the classification performance and thereby effectiveness in using ML models. This aligns with \citet{du2020advances}, who argue that accounting for spatial effects improves both regression and classification performance in PMT models.

The following section outlines applications of ML-PMT models  incorporating geographical factors but do not address spatial correlation effects unlike this study. Details of the data preparation process and the methodological framework are provided in Section \ref{sec:datamethods}, followed by the results of implementing the SML-PMT model using DTKS data from Indonesia (Section \ref{sec:Sres}). Section \ref{sec:conc} concludes with a discussion of the findings and recommendations for future research directions.

\section{Machine learning and geographical effects in PMT and CBT}

The identification of beneficiaries for social assistance programs remains a significant challenge in developing countries. Traditional methods such as Proxy Means Testing (PMT) and Community-Based Targeting (CBT) are widely used for poverty targeting. However, these approaches often exhibit biases that can exacerbate inequities, leading to the systematic exclusion of vulnerable groups \citep{dietrich2024}. For instance, \citet{schnitzer2024} compares PMT and CBT across nine programs in six Sahelian countries. While PMT outperforms CBT in identifying poor households, both methods show limited effectiveness in homogeneous, low-income settings. 

\citet{mcbride2018} also emphasize the importance of out-of-sample validation in improving the predictive accuracy of PMT tools, with and without geographical effects. By employing stochastic ensemble methods and cross-validation, \citet{mcbride2018} achieves significant improvements in PMT performance. However, their model does not explicitly account for geographic variations, limiting the robustness of PMT tools across diverse regions.\citet{schnitzer2024} emphasizes the potential of geographic targeting at regional or village levels in improving PMT performance. 

Recent advancements in geospatial analysis and machine learning (ML) offer a promising approach in integrating geographic variability with household data. \citet{poulin2022} explores an ML-based Proxy Means Test (ML-PMT) for identifying the poorest households in rural Ghana focusing exclusively on water subsidies. The ML-PMT outperforms traditional methods, including the Poverty Probability Index (PPI), CBT, and Ghana's Livelihood Empowerment Against Poverty (LEAP) program, by achieving higher accuracy in excluding wealthier households. This reduces biases which undermine the effectiveness of social programs. Similarly, \citet{solis2022} introduces a machine learning framework utilizing XGBoost to predict poverty in Costa Rica, improving upon traditional methods like PMT. 

\citet{solis2022}'s model, which incorporates occupation and appliance indicators, strikes a better balance between inclusion and exclusion errors, demonstrating the potential of ML to enhance the precision and cost-effectiveness of social assistance targeting.  \citet{gualavisi2024} combines household consumption surveys with geospatial data to target poor villages in Malawi, using XGBoost and satellite data. Their results indicate that incorporating geographic indicators improves prediction accuracy for village welfare, with a rank correlation of 0.75, compared to traditional PMT methods which show lower correlations ($r$ ranging from -0.02 to 0.2). However, while promising, this approach is limited by the lack of out-of-sample validation and consideration of exclusion and inclusion errors. 

Recently, \citet{nasri2024} also introduced two innovative methods—Mixed Means Test (MMT) and multidimensional targeting—for social safety nets in Tunisia. MMT extends PMT by integrating geographic variables, reducing inclusion and exclusion errors. \citet{leite2014} further supports the combination of PMT with geographical and community-based targeting, asserting that this integrated approach can improve efficiency and effectiveness by addressing the limitations inherent in individual methods. 

The benefits of integrating geographical targeting are also evident in \citet{schnitzer2019}’s study on adaptive social protection in Niger, where PMT, Household Economy Analysis (HEA), and geographic targeting methods are compared with each other. While PMT is more effective for identifying persistently poor households, HEA is better for transiently food-insecure households. The study recommends combining these methods based on program goals to maximize efficiency. Similarly, \citet{fortin2016} evaluates a food voucher program in Burkina Faso, showing that while geographical targeting effectively identifies vulnerable districts, PMT struggles to select the most vulnerable households within those districts. This highlights the need for improved methods that combine geographical targeting with household-level data to enhance the identification of vulnerable groups.

Such studies, as highlighted in this paper so far, suggest the critical role of integrating geographic information into poverty prediction models. However, several significant gaps remain in the literature. First, many studies are constrained by the inclusion of geographic information without explicitly addressing spatial correlation between regions, a key focus of our research. Second, the exclusion of poor households, an essential dimension of the precision of the target, is not explored enough in most existing studies.

Addressing these gaps by incorporating spatial contiguity and conducting cross-validations of inclusion and exclusion errors has the potential to significantly enhance the performance of poverty-targeting models. In this research, we propose a hybrid model that integrates Proxy Means Testing (PMT) with spatial contiguity matrices at sub-national levels to improve both the accuracy and equity of social assistance programs. By bridging the divide between household-level socioeconomic data and regional disparities, our SML-PMT model offers a more comprehensive and effective approach to poverty alleviation.
\section{Data and Methods}\label{sec:datamethods}

\subsection{Data}

Two datasets are used in this study: (1) data of SUSENAS from 2016 to 2020, and (2) data of SUSENAS from 2016 to 2021. The data for 2016 to 2020 includes n = 1,533,746 entries and 134 variables. The data for 2016 to 2021 includes n = 1,512,887 entries and 255 variables. The data between 2016 to 2021 includes variables not present in the dataset of the years 2016 to 2020, such as indexes of access to basic services, infrastructure, as well as binomial and multinomial variables that measure access to essential locations such as hospitals due to 2021's questionnaire being modified to take into consideration the COVID-19 pandemci at the time. Despite a deeper dive into access and deprivation of households, a large number of missing values exist in 2021 which is reflected in the updated dataset. 

In both datasets, the dependent (target) variable $y$ is per capita expenditure. The target variable $y$ is a numerical (continuous) variable, but it is transformed into a binary variable equal to 1 for poor and 0 for non-poor households. This transformation is applied with a 40\% poverty threshold, as in the Data Terpadu Kesejahteran Sosial (DTKS) scheme of social protection. 

\subsubsection{Data preparation}

The categorical explanatory variables were transformed into numerical variables through one-hot encoding. For example, in the case of gender, the categories male and female are consolidated into a single numeric variable equal to 1 for female and equal to 0 for male. For categorical variables with no previous ordinal relationship, different binary variables are generated for each category. For example, the variable economic sector has 3 categories: agriculture, industry, and services. This categorical variable is converted to 3 binary variables: a variable of agricultural sector equal to 1 for agriculture and 0 for industry and services, a variable of industry equal 1 for industry and 0 for agriculture and services, and a variable of services equal to 1 for services and 0 for agriculture and industry. 
 
\begin{figure}[ht]
\centering
\captionsetup{justification=centering,margin=1cm}
\caption{Outlier treatment: target variable (logarithm of per capita expenditure). Left: original variable; right: treated variable.}
\includegraphics[width=5in]{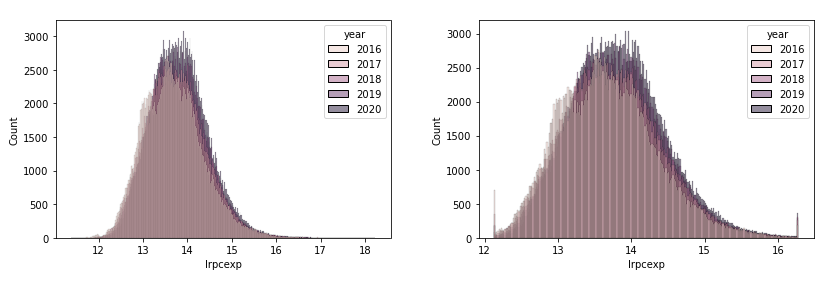}
\label{fig:pce_win}
\end{figure}

\begin{figure}[ht]
\centering
\captionsetup{justification=centering,margin=1cm}
\caption{Outlier treatment of the explanatory variables (ikg = village difficulty index). Left: original variable; right: treated variable.}
\includegraphics[width=5in]{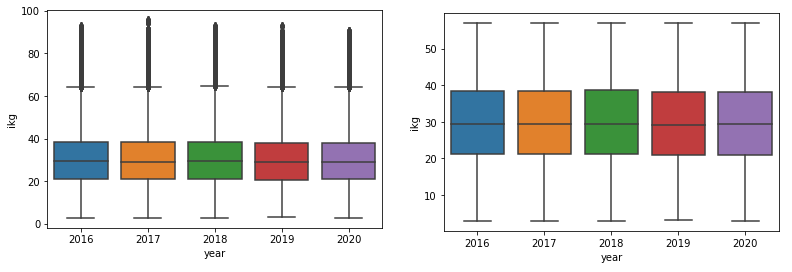}
\label{fig:ikg_win}
\end{figure}

Outliers were identified and treated by re-coding the extreme values of the variables. Figure \ref{fig:pce_win} shows the outlier treatment of the dependent variable. Figure \ref{fig:ikg_win} shows the treatment of outliers in the village difficulty index. In this case, values of the village difficulty index higher than 60 are re-categorised as `60 and higher' (Figure \ref{fig:ikg_win}, right). 

\begin{figure}[ht]
\centering
\includegraphics[width=5in]{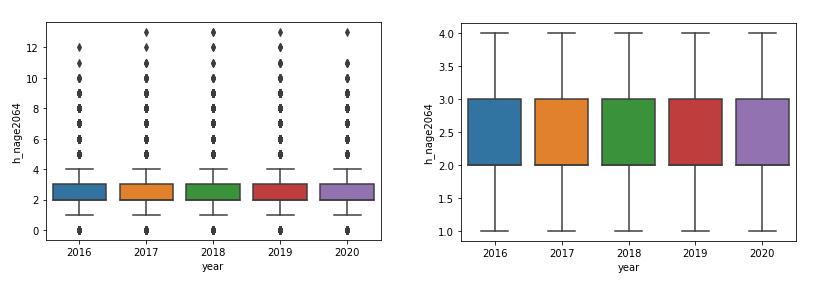}
\caption{Outlier treatment of the explanatory variables (number of people aged 20-64 years in the household) Left: original variable; right: treated variable.}
\label{fig:nage2064_win}
\end{figure}

Figure \ref{fig:nage2064_win} (left) shows that those aged 20-64 years in a household includes outliers both in the lower age category and the upper age category. The households with no individuals between 20-64 years re-categorised as households with 1 or less individuals aged 20-64 years, and those households with more than 4 individuals were re-categorised as households with 4 or more individuals aged 20-64 years.

\FloatBarrier

The missing values for 2021 in the dataset for 2016 to 2021 were imputed with average values in the same kabupaten (district) between 2016 and 2020. This type of imputation assumes a similar trend was observed in the socio-economic indicators of each region (district) in the year 2021, compared to the average of previous years.  

In the spatial machine learning implementation, 80\% of the data was separated in a test sample for validation, and 20\% of the data was used as a train sample to estimate machine learning models. The models fitted in the train sample predict the values of real per capita expenditure in the test sample. 

\begin{figure}[ht]
\centering
\captionsetup{justification=centering,margin=1cm}
\caption{Results of the Principal Component Analysis (PCA) applied to the SUSENAS data 2016 to 2020}
\includegraphics[width=4.2in]{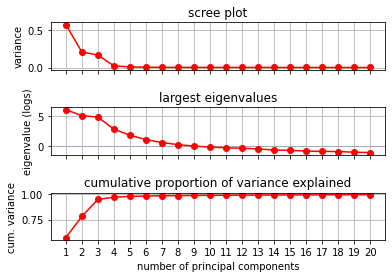}
\label{fig:PCA}
\end{figure}

Principal component analysis (PCA) was applied to summarize a large number of variables into a smaller set of principal components, which are a linear combination of the original variables. Figure \ref{fig:PCA} shows the principal component analysis applied to the SUSENAS data (2016 to 2020). A visual approach suggests to maintain a minimum of 3 components due to the elbow in the scree plot after the third principal component (Figure \ref{fig:PCA}, top). The first 3 components explain more than 90\% of cumulative variance (Figure \ref{fig:PCA}, bottom). The Kaiser rule suggests including at least 10 principal components with eigenvalues greater than 1 (Figure \ref{fig:PCA}, middle). Maximum likelihood was used to choose the optimal number of principal components when PCA is applied to separate regions. See \cite{minka2000automatic} for details on the automatic choice of dimensionality for PCA.

\subsection{Spatial Machine Learning PMT Models}

In Proxy Means Testing (PMT) for poverty targeting, spatial machine learning integrates geographic information to capture spatial dependencies, enabling more accurate predictions of welfare outcomes. Let $\mathbf{X} = [\mathbf{x}_1, \mathbf{x}_2, \dots, \mathbf{x}_n]^\top \in \mathbb{R}^{n \times p}$ denote the matrix of predictors (input features), where $n$ is the number of observations and $p$ is the number of predictors. Let $\mathbf{y} = [y_1, y_2, \dots, y_n]^\top \in \mathbb{R}^n$ represent the response variable, which corresponds to poverty status or welfare measures. A traditional machine learning model is defined as:

\[
\hat{\mathbf{y}} = f(\mathbf{X}; \boldsymbol{\theta}),
\]
where $f$ is a machine learning algorithm (e.g., linear regression, random forest, gradient boosting, or neural network) parameterized by $\boldsymbol{\theta}$.

To explicitly incorporate spatial relationships, the geographical space is divided into $k$ clusters, $\{C_1, C_2, \dots, C_k\}$, based on spatial correlation. These clusters are not obtained using machine learning but are calculated from the spatial contiguity matrix, $\mathbf{W} \in \mathbb{R}^{n \times n}$, which captures the relationships between geographic units. The elements of $\mathbf{W}$ are defined as:

\[
w_{ij} =
\begin{cases}
1 & \text{if locations } i \text{ and } j \text{ are neighbors}, \\
0 & \text{otherwise.}
\end{cases}
\]

The spatial correlation between locations is derived from $\mathbf{W}$, grouping observations into clusters. In this case, the observations within a cluster exhibit high spatial similarity. Using $\mathbf{W}$, the spatially lagged response is defined as:
\[
\mathbf{y}_W = \mathbf{W} \mathbf{y},
\]
which measures the weighted influence of neighboring observations' poverty status on a given observation. Similarly, spatially lagged predictors can be constructed as:
\[
\mathbf{Z} = \mathbf{W} \mathbf{X},
\]
where $\mathbf{Z} \in \mathbb{R}^{n \times p}$ represents the average or weighted aggregate of predictor values from neighboring observations.

In the Spatial Machine Learning (SML) framework, separate machine learning models are estimated for each spatial cluster $C_k$. Within each cluster, let $\mathbf{X}_k$, $\mathbf{y}_k$, and $\mathbf{Z}_k$ represent the predictors, response, and spatially augmented features for cluster $k$, respectively. The model for cluster $k$ can be expressed as:

\[
\hat{\mathbf{y}}_k = f_k(\mathbf{X}_k, \mathbf{Z}_k; \boldsymbol{\theta}_k),
\]
where $f_k$ is the machine learning model specific to cluster $k$, and $\boldsymbol{\theta}_k$ are the parameters learned for this cluster. The cluster-specific predictions are then combined to generate the final global predictions:
\[
\hat{\mathbf{y}} = \bigcup_{k=1}^K \hat{\mathbf{y}}_k,
\]
where $\bigcup$ represents the union of predictions from all clusters. This approach ensures that the models adapt to the unique characteristics of each cluster while explicitly accounting for the spatial dependencies derived from the contiguity matrix $\mathbf{W}$. The integration of spatial features ($\mathbf{Z}$) enhances the accuracy and equity of poverty targeting by leveraging the spatially correlated structure of the data.

\subsubsection{Model evaluation}

The precision of the methods is evaluated with inclusion and exclusion errors. Let $tp$ be the number of poor households that are correctly classified as poor by machine learning algorithms. These correctly classified cases are called \textit{true positives} ($tp$). Let $tn$ be the non-poor households, which are also correctly classified as non-poor by the machine learning algorithms. These are called \textit{true negatives} ($tn$). Let $fp$ be the non-poor households in the data set, which are incorrectly classified as poor; these are defined as \textit{false positives} ($fp$). Finally, let $fn$ be the poor households, which are incorrectly classified as non-poor. These individuals are \textit{false negatives} ($fn$). Based on these definitions, exclusion errors (EE) and inclusion errors (IE) are calculated with the formulas below. 
\[
\text{EE = } \frac{\text{fn}}{\text{tp + fn}}
\]
\[
\text{IE = } \frac{\text{fp}}{\text{tp + fp}}
\]
These are the formulas used by \cite{noriega2020algorithmic} for algorithmic targeting. In terms of targeting beneficiaries, exclusion errors represent the proportion of poor households which may have been wrongfully excluded from the program. Inclusion errors indicate the proportion of non-poor households that were wrongfully included in the program. Besides inclusion and exclusion errors, sensitivity, specificity and the R squared ($\text{R}^\text{2}$) metrics are calculated to evaluate the performance of the ML models in the validation (test) sample.

\section{Results}\label{sec:Sres}

\subsection{Spatial contiguity matrix}

\begin{figure}[ht]
\centering
\captionsetup{justification=centering,margin=1cm}
\caption{Spatial contiguity by region (kabupaten)}
\includegraphics[width=5in]{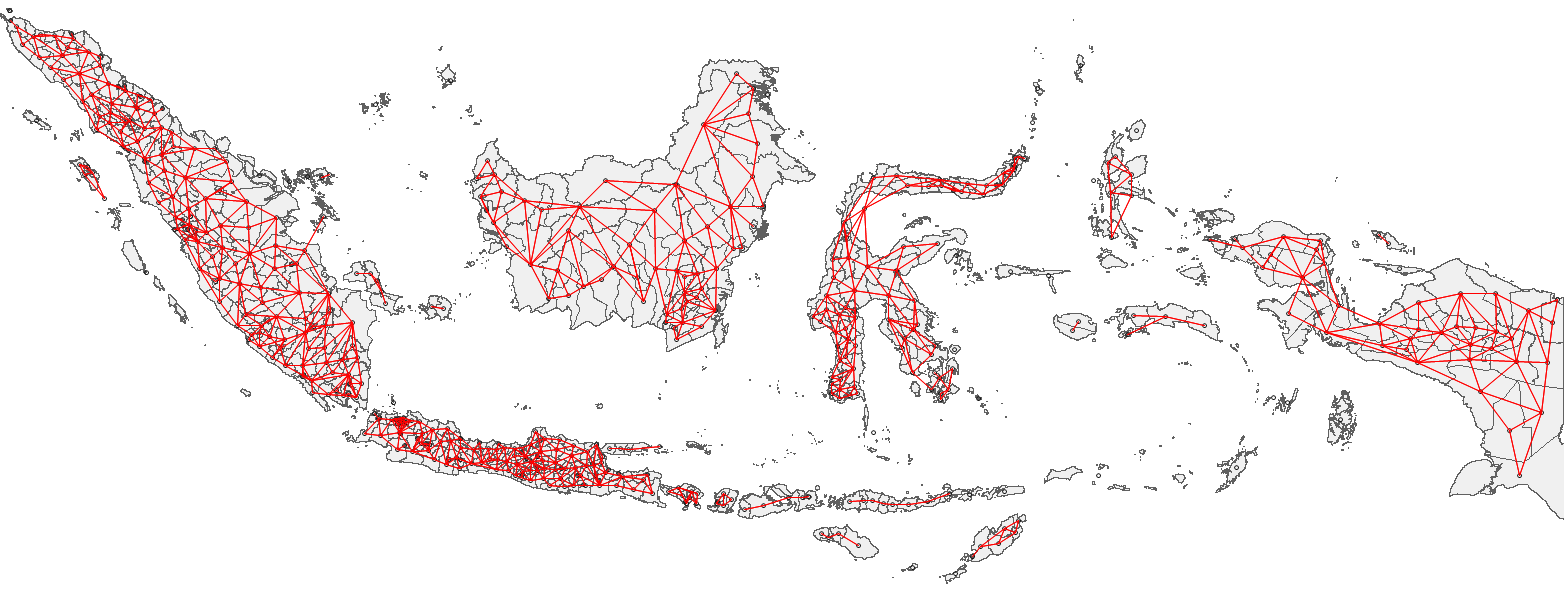}
\label{fig:matW}
\end{figure}

The spatial contiguity matrix (Figure \ref{fig:matW}) was calculated based on the centroids of each region (kabupaten) obtained from Delaunay Triangulation. Delaunay Triangulation calculates the neighbours of regions by creating Voronoi triangles from the centroids of the kabupaten such that each centroid is a triangle node. Nodes connected by a triangle edge are considered neighbours to one another. Using Delaunay triangulation allows every feature to have at least one neighbour even when the data includes islands or widely varying feature densities as in the case of this data collected in Indonesia. 

\subsection{Spatial patterns in Indonesia}

\begin{figure}[ht]
\centering
\captionsetup{justification=centering,margin=1cm}
\caption{Distribution of average per capita expenditure by region (Kabupaten)}
\includegraphics[width=5in]{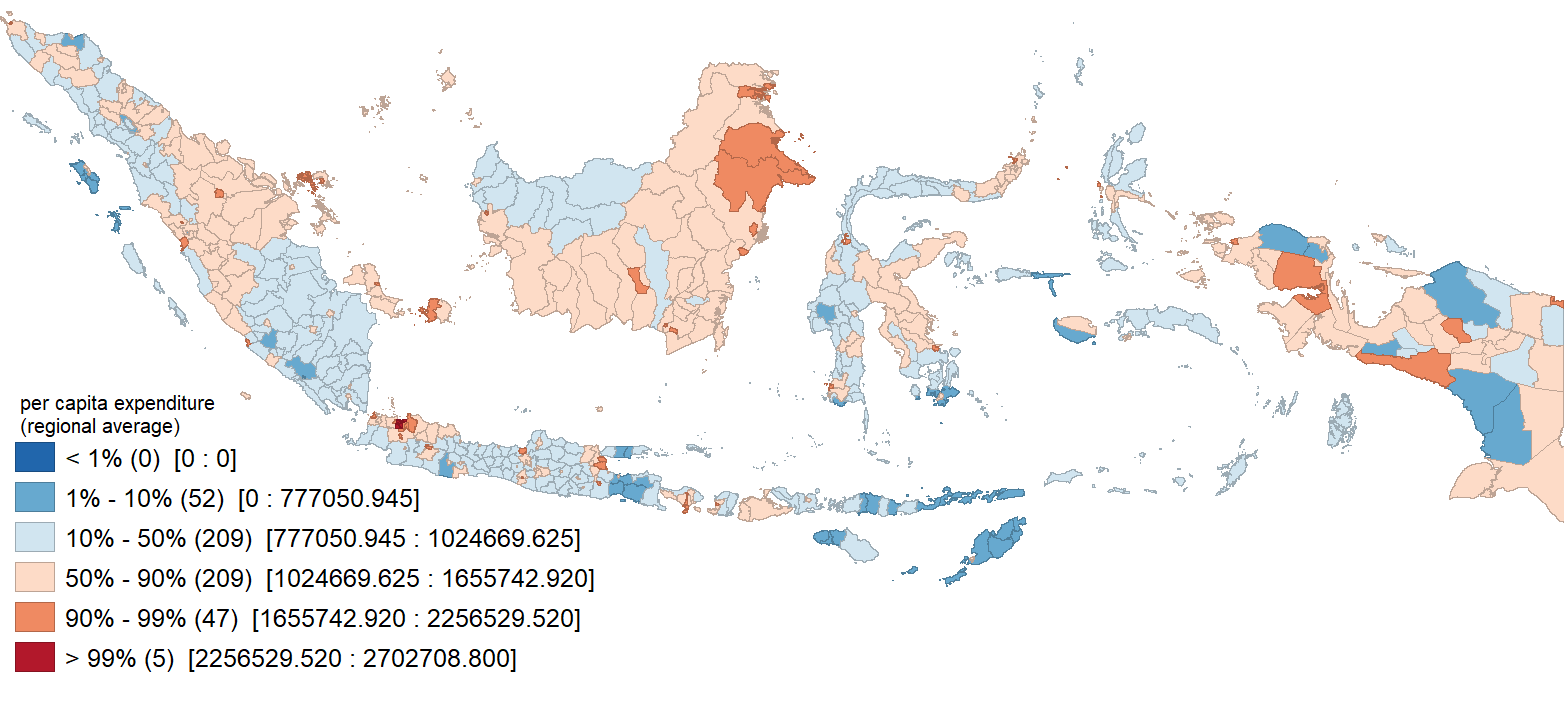}
\label{fig:apcexp}
\end{figure}

The average per capita expenditure of Indonesian households were influenced by their geography. Figure \ref{fig:apcexp} shows the average per capita expenditure by region between 2016 and 2020. A clear geographical) pattern can be observed in the map as per capita expenditure rises above the 50th percentile. This is particularly clear in North Kalimantan and East Kalimantan, where in some regions, households have an average per capita expenditure above the 90th percentile. In contrast, lower expenditure can be seen for households in Lampung, South Sumatra, and Bengkulu. In the province of Papua and West Papua, there is a clear heterogeneity of per capita expenditure, as some regions in these provinces have very low per capita expenditure (below the 10th percentile) while other spend above the 90th percentile on a per capita basis (Figure \ref{fig:apcexp}). 

\begin{figure}[ht]
\centering
\captionsetup{justification=centering,margin=1cm}
\caption{Moran's I statistic for the spatial autocorrelation of average per capita expenditure between regions (Kabupaten)}
\includegraphics[width=5in]{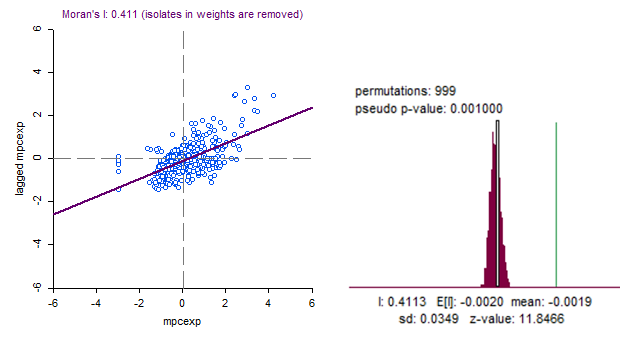}
\label{fig:moran}
\end{figure}

The existence of the spatial correlation between the per capita expenditure of households in the regions of Indonesia is confirmed by Moran's I statistic (Figure \ref{fig:moran}). Moran's I is a measure of overall spatial correlation calculated as the cross-product between average per capita expenditure and its spatial lag, with the variable expressed in deviations from its mean and weighted by the spatial contiguity matrix. The estimated value of Moran's I statistic for Indonesia is equal to 0.411 and has an associated significance level of less than 1\% (p-value = 0.001). The null hypothesis of spatial randomness can be rejected at conventional significance levels based on a permutation test (Figure \ref{fig:moran}, right). 

\begin{figure}[ht]
\centering
\captionsetup{justification=centering,margin=1cm}
\caption{Moran's I statistic for the spatial auto-correlation of average per capita expenditure between high-expenditure regions (Kabupaten)}
\includegraphics[width=5in]{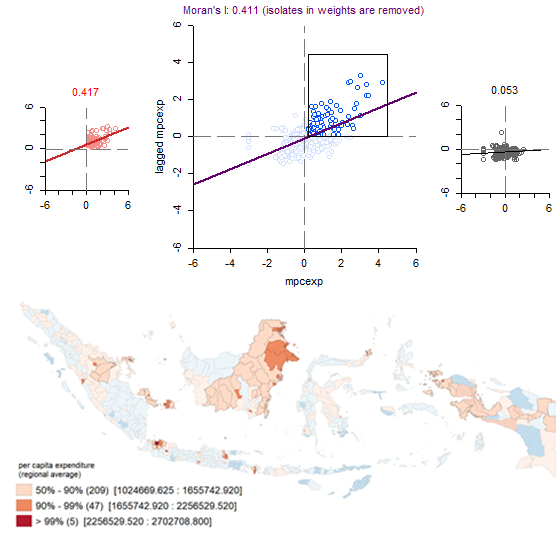}
\label{fig:moran1}
\end{figure}

\begin{figure}[ht]
\centering
\captionsetup{justification=centering,margin=1cm}
\caption{Moran's I statistic for the spatial auto-correlation of average per capita expenditure between low-expenditure regions (Kabupaten)}
\includegraphics[width=5in]{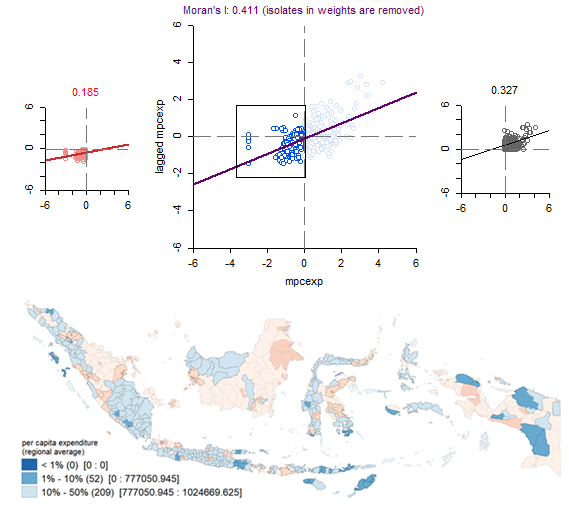}
\label{fig:moran2}
\end{figure}

Figures \ref{fig:moran1} and \ref{fig:moran2} show the regionalized value of Moran's I calculated for a subset of regions with high per capita expenditure (Figure \ref{fig:moran1}) and for a subset of regions with low per capita expenditure (Figure \ref{fig:moran2}). A higher spatial correlation is observed for regions with high per capita expenditure (Moran's I = 0.417), compared to regions with low per capita expenditure (Moran's I = 0.185). To capture this heterogeneity in the local spatial dependence, the Getirs-Ord statistic \citep{getis2010analysis} was calculated for the per capita expenditure of households in all the Kabupaten or regions of Indonesia. 

\begin{figure}[ht]
\centering
\captionsetup{justification=centering,margin=1cm}
\caption{Local spatial auto-correlation (Getis-Ord statistic) of per capita expenditure by regions (Kabupaten)}
\includegraphics[width=5in]{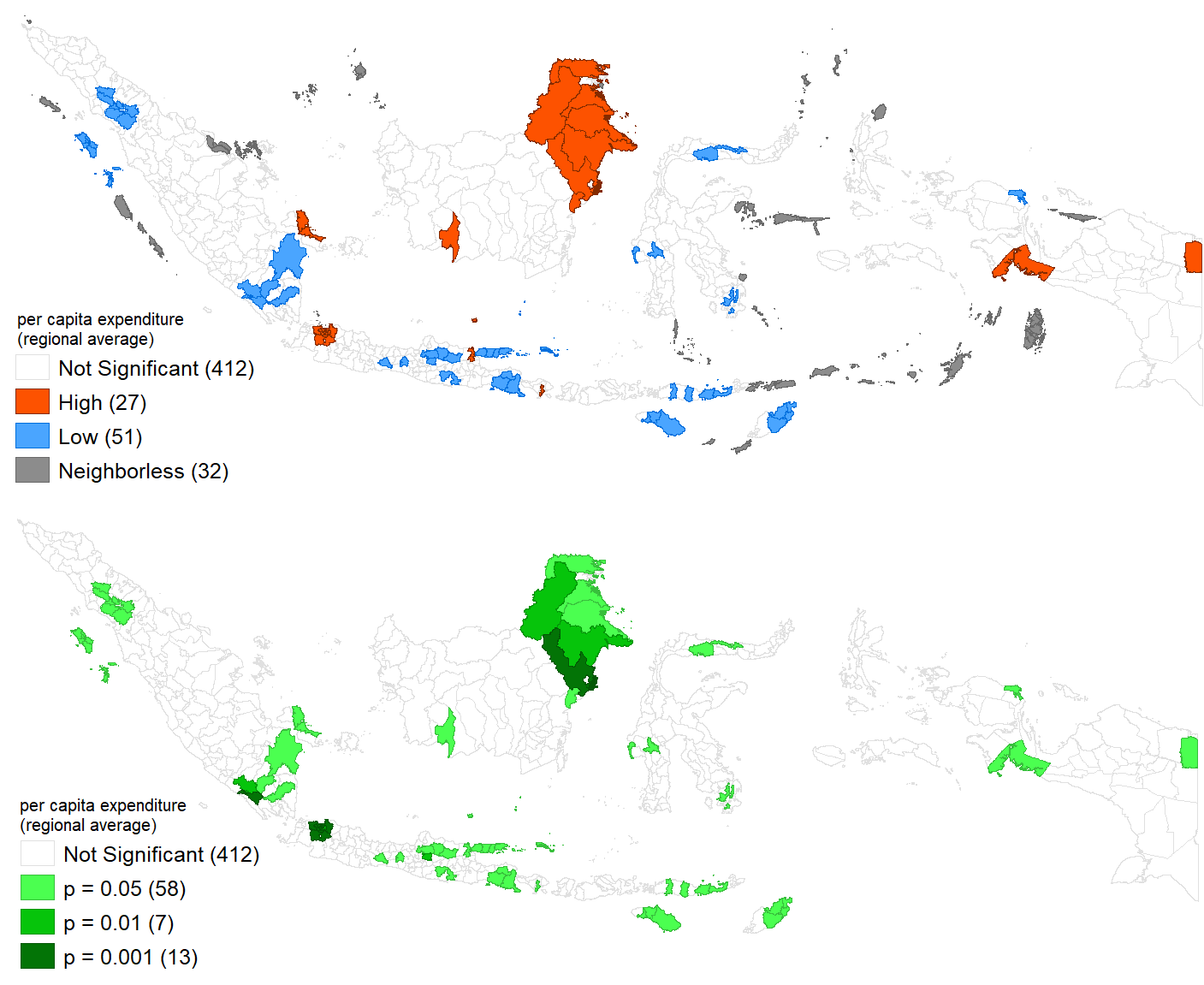}
\label{fig:GOclusters}
\end{figure}

The results show a significant local spatial correlation for households with high per capita expenditure ub North Kalimantan and East Kalimantan (p-value < 0.05), and between the regions of West Papua. Jakartan households also showed significant spatial correlation in terms of high per capita expenditure (Figure \ref{fig:GOclusters}). In turn, households in Gorontalo,East Nusa Tenggara, South Sumatra, North Sumatra, Central Java and East Java had significant local spatial correlations in terms of low per capita expenditure. These spatial patterns will be included in refining the machine learning algorithms further in an effort to effectively target exclusion and inclusion errors.

\subsection{Spatial Hierarchical Clustering}

Spatial hierarchical clustering imposes contiguity as a constraint in a clustering procedure. This clustering groups an original set of $n$ spatial units (kabupaten) into $p$ internally connected regions that maximise within similarity. Contiguity is a constraint in the grouping because when using the centroids of each region, the groups consisting of kabupaten are geographically connected. The clustering is hierarchical, because it starts with the full set of kabupaten as one cluster, and finds the best break point to create two clusters. This process continues until each observation is its own cluster. The results of the successive divisions of the data is visualized in a graphic called a dendrogram, which has a representation similar to a tree. In the dendogram, the dashed red line corresponds to a cut point that yields a specific number of clusters.

\begin{figure}[ht]
\centering
\caption{Spatial hierarchical clustering and dendogram: 4 clusters}
\includegraphics[width=5in]{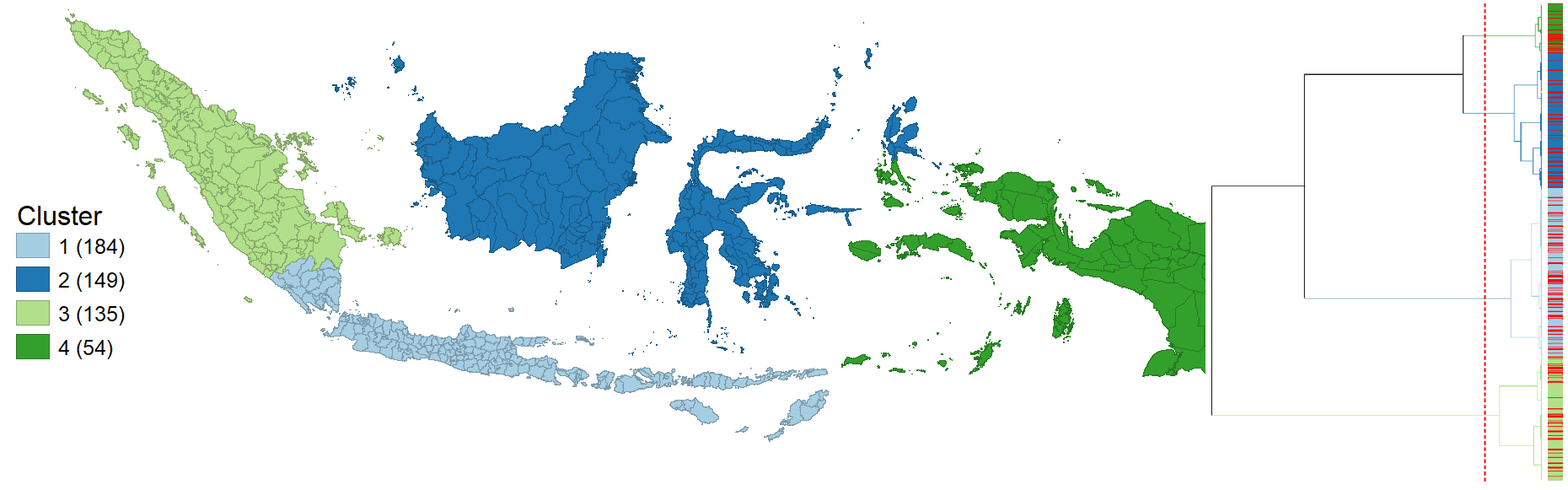}
\label{fig:hclus4}
\end{figure}

\begin{figure}[ht]
\centering
\caption{Spatial  hierarchical clustering and dendogram: 6 clusters}
\includegraphics[width=5in]{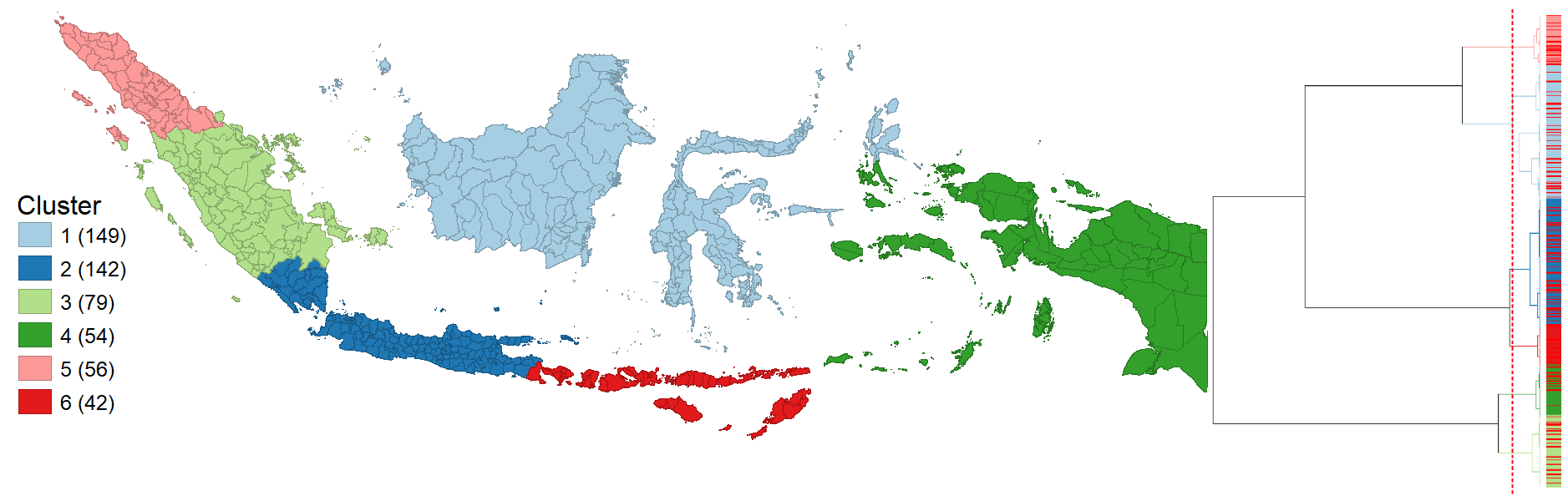}
\label{fig:hclus6}
\end{figure}

\begin{figure}[ht]
\centering
\caption{Spatial  hierarchical clustering and dendogram: 12 clusters}
\includegraphics[width=5in]{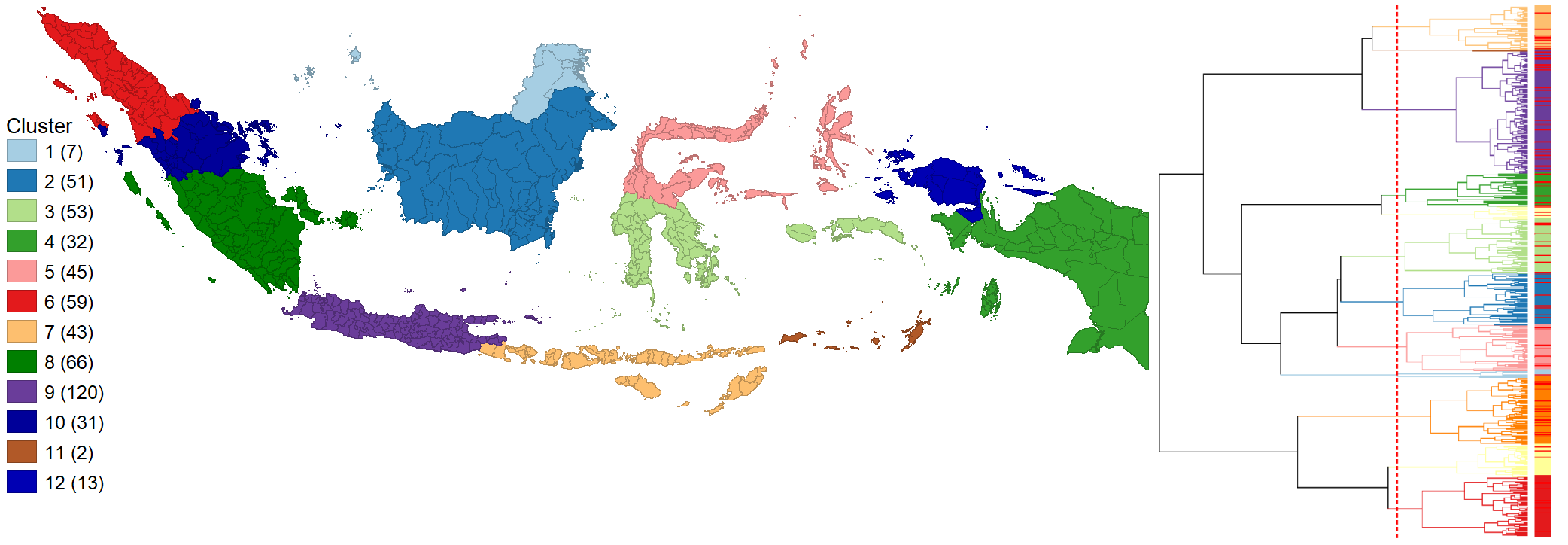}
\label{fig:hclus12}
\end{figure}

\subsection{Spatial machine learning results}

\begin{figure}[ht]
\centering
\captionsetup{justification=centering,margin=1cm}
\caption{Machine learning results without spatial component (benchmark models): exclusion errors}
\includegraphics[width=5in]{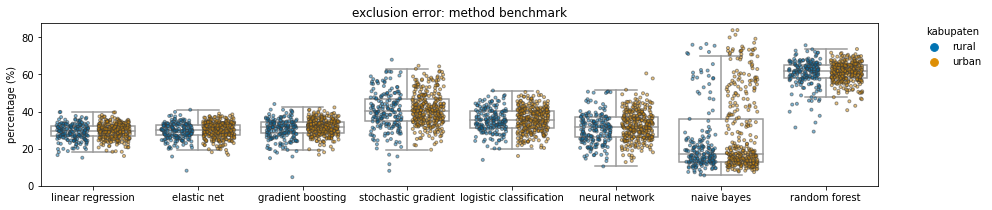}
\label{fig:ee_ben}
\end{figure}

Table \ref{tab:smlres} shows the performance of the spatial machine learning models with and without principal component analysis for different geographical clusters in Indonesia obtained with spatial hierarchical clustering applied to the data from 2016 to 2020. Table \ref{tab:smlres_21} shows the performance of the spatial machine learning models again with and without principal component analysis, and for different geographical clusters, obtained with the updated and imputed data from 2016 to 2021. In both tables, the average exclusion error and inclusion error of the spatial machine learning are compared against a benchmark machine learning model without spatial components (Figures \ref{fig:ee_ben} and \ref{fig:ee_ben}). The sn method (separate spatially; do not separate/normalize temporally) is considered the most effective model (the model with the lowest average exclusion error and inclusion error) in the latest machine learning PMT implementation without spatial analysis. 

\begin{figure}[ht]
\centering
\captionsetup{justification=centering,margin=1cm}
\caption{Spatial machine learning results: exclusion errors, 4 clusters (without principal component analysis)}
\includegraphics[width=5in]{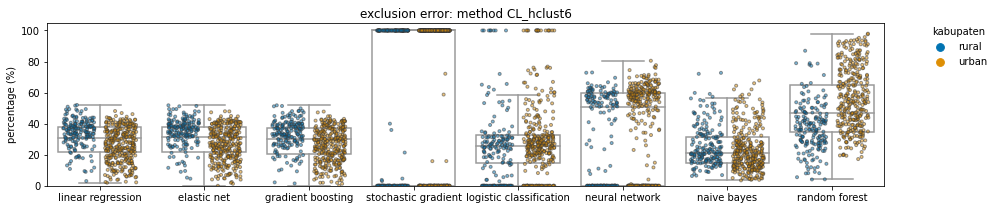}
\label{fig:eenpc_hclus4}
\end{figure}

\begin{figure}[ht]
\centering
\captionsetup{justification=centering,margin=1cm}
\caption{Spatial machine learning results: inclusion errors, 4 clusters (without principal component analysis)}
\includegraphics[width=5in]{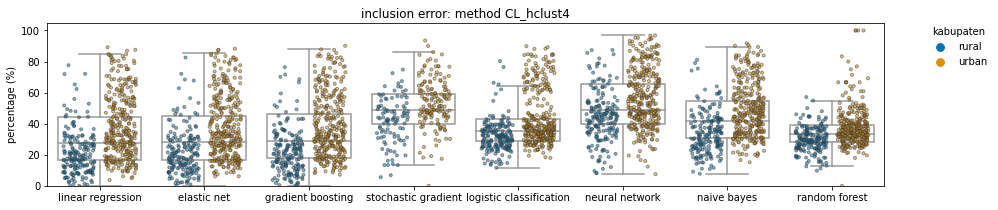}
\label{fig:ienpc_hclus4}
\end{figure}

\begin{figure}[ht]
\centering
\captionsetup{justification=centering,margin=1cm}
\caption{Spatial machine learning results: exclusion errors, 6 clusters (without principal component analysis)}
\includegraphics[width=5in]{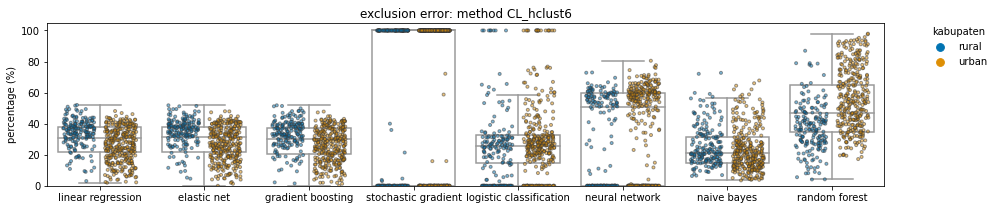}
\label{fig:eenpc_hclus6}
\end{figure}

\begin{figure}[ht]
\centering
\captionsetup{justification=centering,margin=1cm}
\caption{Spatial machine learning results: inclusion errors, 6 clusters (without principal component analysis)}
\includegraphics[width=5in]{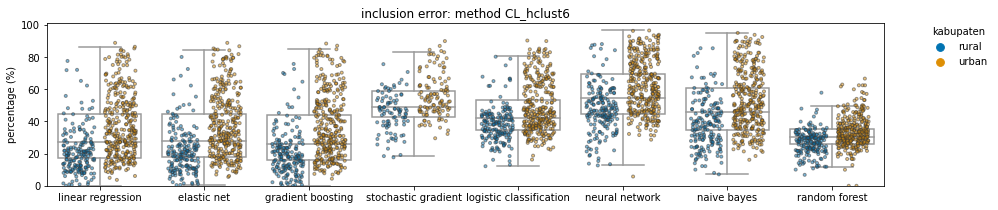}
\label{fig:ienpc_hclus6}
\end{figure}

\begin{figure}[ht]
\centering
\captionsetup{justification=centering,margin=1cm}
\caption{Spatial machine learning results: exclusion errors, 12 clusters (without principal component analysis)}
\includegraphics[width=5in]{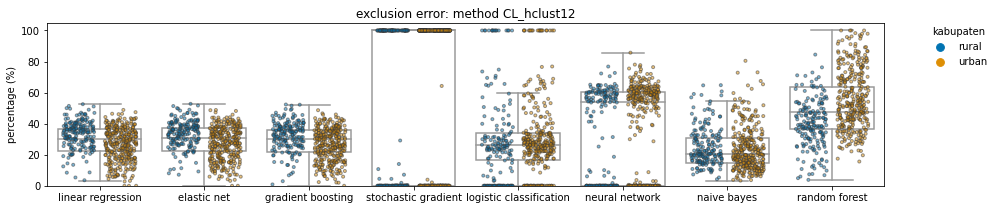}
\label{fig:eenpc_hclus12}
\end{figure}

\begin{figure}[ht]
\centering
\captionsetup{justification=centering,margin=1cm}
\caption{Spatial machine learning results: inclusion errors, 12 clusters (without principal component analysis)}
\includegraphics[width=5in]{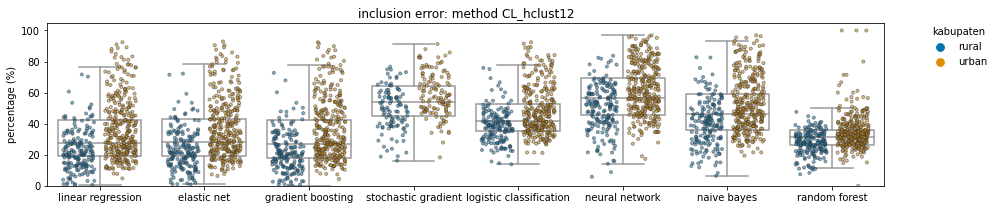}
\label{fig:ienpc_hclus12}
\end{figure}

\begin{figure}[ht]
\centering
\captionsetup{justification=centering,margin=1cm}
\caption{Spatial machine learning results: exclusion errors, 4 clusters (with principal components)}
\includegraphics[width=5in]{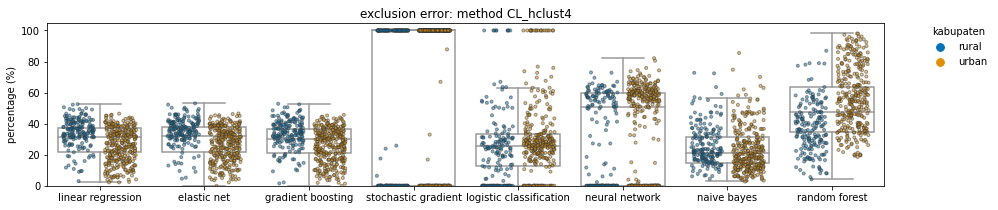}
\label{fig:ee_hclus4}
\end{figure}

\begin{figure}[ht]
\centering
\captionsetup{justification=centering,margin=1cm}
\caption{Spatial machine learning results: inclusion errors, 4 clusters (with principal components)}
\includegraphics[width=5in]{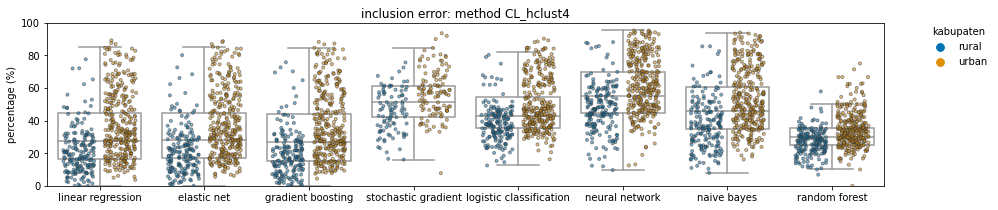}
\label{fig:ie_hclus4}
\end{figure}

\begin{figure}[ht]
\centering
\captionsetup{justification=centering,margin=1cm}
\caption{Spatial machine learning results: exclusion errors, 6 clusters (with principal components)}
\includegraphics[width=5in]{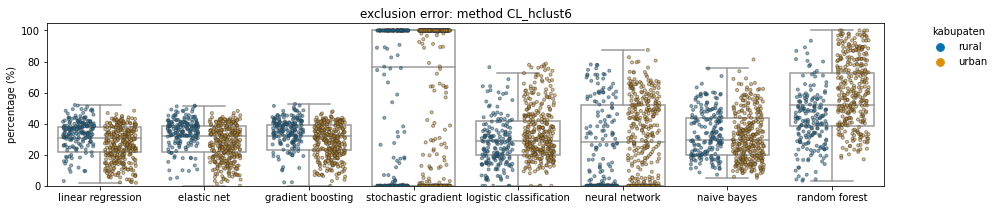}
\label{fig:ee_hclus6}
\end{figure}

\begin{figure}[ht]
\centering
\captionsetup{justification=centering,margin=1cm}
\caption{Spatial machine learning results: inclusion errors, 6 clusters (with principal components)}
\includegraphics[width=5in]{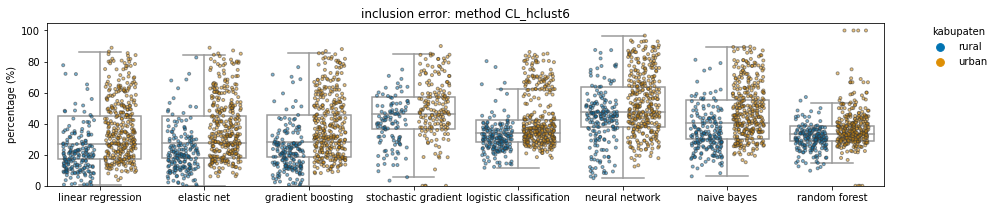}
\label{fig:ie_hclus6}
\end{figure}

\begin{figure}[ht]
\centering
\captionsetup{justification=centering,margin=1cm}
\caption{Spatial machine learning results: exclusion errors, 12 clusters (with principal components)}
\includegraphics[width=5in]{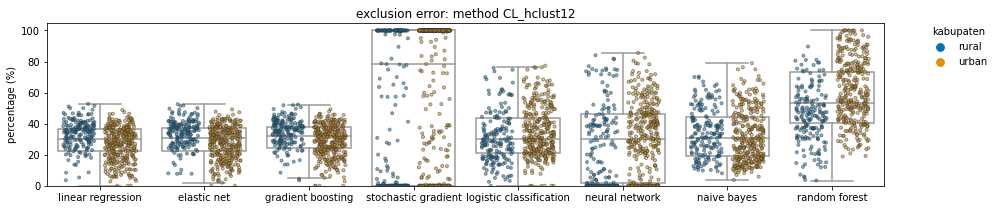}
\label{fig:ee_hclus12}
\end{figure}

\begin{figure}[ht]
\centering
\captionsetup{justification=centering,margin=1cm}
\caption{Spatial machine learning results: inclusion errors, 12 clusters (with principal components)}
\includegraphics[width=5in]{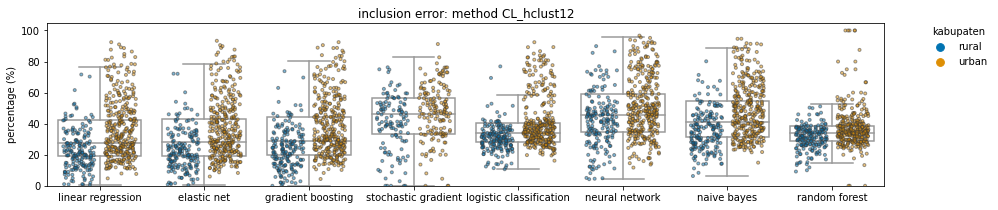}
\label{fig:ie_hclus12}
\end{figure}

In the data from 2016 to 2020 (Table \ref{tab:smlres}), the spatial contiguity exercise reduces the average exclusion error in the test sample from 28\% to 20\%. The lowest EE (28\%) in the machine learning implementation without a spatial component is obtained with a linear regression model when estimating the data at the standard 40 per cent poverty threshold. With spatial machine learning models, the lowest average exclusion error (equal to 20\%) is obtained with a machine learning model based on a Naive Bayes classifier, without PCA. Similar results are obtained when using 4 spatial clusters (Figures \ref{fig:eenpc_hclus4} and \ref{fig:ienpc_hclus4}), 6 spatial clusters (Figures \ref{fig:eenpc_hclus6} and \ref{fig:ienpc_hclus6}), or 12 spatial clusters (Figures \ref{fig:eenpc_hclus12} and \ref{fig:ienpc_hclus12}). With PCA, the lowest exclusion error (23\%) is also obtained with a Naive Bayes classifier. Implementing PCA led to the results being inordinately dispersed across all the models for both, inclusion errors and exclusion errors. see Figures \ref{fig:ee_hclus4}, \ref{fig:ie_hclus4}, \ref{fig:ee_hclus6}, \ref{fig:ie_hclus6}, \ref{fig:ee_hclus12}, \ref{fig:ie_hclus12}. 

The dataset from 2016 to 2021 (Table \ref{tab:smlres_21}) shows that the spatial machine learning reduced the average EE from 27\% to 24\% (SML). The lowest EE (24\%) is obtained with a ML model based on a Naive Bayes classifier without PCA. A higher IE on average is obtained with spatial patterns compared to models without spatial patterns. At the provincial level, exclusion errors are lower in urban kabupaten compared to rural kabupaten. The provinces of Nusa Tenggara Timur and Bengkulu have IE below 25\% (Appendix 2).

\begin{landscape}
\begin{table}[htbp]
\small
  \centering
  \captionsetup{justification=centering,margin=1cm}
  \caption{Machine learning results with and without spatial analysis, and with and without principal component analysis: exclusion errors and inclusion errors in the test sample. Data from 2016 to 2020}
    \begin{tabular}{lllllllllll}
    \toprule
    \multicolumn{2}{c}{\multirow{3}[6]{*}{}} & \multicolumn{1}{c}{\multirow{3}[6]{*}{\textbf{\makecell{Benchmark \\ ML models*}}}} &       & \multicolumn{7}{c}{\textbf{Spatial machine learning models}} \\
\cmidrule{5-11}    \multicolumn{2}{c}{} &       &       & \multicolumn{3}{c}{\textbf{Without principal component analysis}} &       & \multicolumn{3}{c}{\textbf{With principal component analysis}} \\
\cmidrule{5-7}\cmidrule{9-11}    \multicolumn{2}{c}{} &       &       & \multicolumn{1}{c}{\textbf{4 clusters}} & \multicolumn{1}{c}{\textbf{6 custers}} & \multicolumn{1}{c}{\textbf{12 clusters}} &       & \multicolumn{1}{c}{\textbf{4 clusters}} & \multicolumn{1}{c}{\textbf{6 custers}} & \multicolumn{1}{c}{\textbf{12 clusters}} \\
    \midrule
    \multicolumn{11}{l}{\textbf{Exclusion error}} \\
          & Elastic net & \multicolumn{1}{c}{\cellcolor[rgb]{ .878,  .886,  .51}28.50} &       & \multicolumn{1}{c}{\cellcolor[rgb]{ .875,  .882,  .51}28.42} & \multicolumn{1}{c}{\cellcolor[rgb]{ .871,  .882,  .51}28.39} & \multicolumn{1}{c}{\cellcolor[rgb]{ .867,  .882,  .51}28.28} &       & \multicolumn{1}{c}{\cellcolor[rgb]{ .875,  .886,  .51}28.47} & \multicolumn{1}{c}{\cellcolor[rgb]{ .875,  .882,  .51}28.46} & \multicolumn{1}{c}{\cellcolor[rgb]{ .871,  .882,  .51}28.38} \\
          & Gradient boosting & \multicolumn{1}{c}{\cellcolor[rgb]{ 1,  .906,  .518}31.48} &       & \multicolumn{1}{c}{\cellcolor[rgb]{ .894,  .89,  .51}28.79} & \multicolumn{1}{c}{\cellcolor[rgb]{ .89,  .89,  .51}28.74} & \multicolumn{1}{c}{\cellcolor[rgb]{ .878,  .886,  .51}28.53} &       & \multicolumn{1}{c}{\cellcolor[rgb]{ 1,  .914,  .518}30.99} & \multicolumn{1}{c}{\cellcolor[rgb]{ 1,  .918,  .518}30.97} & \multicolumn{1}{c}{\cellcolor[rgb]{ 1,  .918,  .518}30.93} \\
          & Linear regression & \multicolumn{1}{c}{\cellcolor[rgb]{ .859,  .878,  .506}28.20} &       & \multicolumn{1}{c}{\cellcolor[rgb]{ .859,  .878,  .506}28.20} & \multicolumn{1}{c}{\cellcolor[rgb]{ .863,  .878,  .506}28.22} & \multicolumn{1}{c}{\cellcolor[rgb]{ .859,  .878,  .506}28.15} &       & \multicolumn{1}{c}{\cellcolor[rgb]{ .863,  .878,  .506}28.22} & \multicolumn{1}{c}{\cellcolor[rgb]{ .863,  .882,  .51}28.25} & \multicolumn{1}{c}{\cellcolor[rgb]{ .859,  .878,  .506}28.19} \\
          & Logistic classification & \multicolumn{1}{c}{\cellcolor[rgb]{ 1,  .875,  .51}33.25} &       & \multicolumn{1}{c}{\cellcolor[rgb]{ .82,  .867,  .506}27.53} & \multicolumn{1}{c}{\cellcolor[rgb]{ .816,  .867,  .506}27.44} & \multicolumn{1}{c}{\cellcolor[rgb]{ .996,  .918,  .514}30.49} &       & \multicolumn{1}{c}{\cellcolor[rgb]{ 1,  .875,  .51}33.25} & \multicolumn{1}{c}{\cellcolor[rgb]{ 1,  .878,  .51}33.15} & \multicolumn{1}{c}{\cellcolor[rgb]{ 1,  .863,  .51}33.93} \\
          & Naive Bayes & \multicolumn{1}{c}{\cellcolor[rgb]{ .894,  .89,  .51}28.77} &       & \multicolumn{1}{c}{\cellcolor[rgb]{ .404,  .749,  .482}20.44} & \multicolumn{1}{c}{\cellcolor[rgb]{ .4,  .749,  .482}20.40} & \multicolumn{1}{c}{\cellcolor[rgb]{ .388,  .745,  .482}20.14} &       & \multicolumn{1}{c}{\cellcolor[rgb]{ .549,  .788,  .49}22.90} & \multicolumn{1}{c}{\cellcolor[rgb]{ .565,  .796,  .49}23.17} & \multicolumn{1}{c}{\cellcolor[rgb]{ .557,  .792,  .49}23.05} \\
          & Neural network & \multicolumn{1}{c}{\cellcolor[rgb]{ 1,  .922,  .518}30.60} &       & \multicolumn{1}{c}{\cellcolor[rgb]{ 1,  .859,  .506}34.19} & \multicolumn{1}{c}{\cellcolor[rgb]{ .996,  .839,  .502}35.23} & \multicolumn{1}{c}{\cellcolor[rgb]{ .996,  .831,  .502}35.69} &       & \multicolumn{1}{c}{\cellcolor[rgb]{ 1,  .894,  .514}32.08} & \multicolumn{1}{c}{\cellcolor[rgb]{ .945,  .906,  .514}29.66} & \multicolumn{1}{c}{\cellcolor[rgb]{ 1,  .902,  .514}31.77} \\
          & Random forest & \multicolumn{1}{c}{\cellcolor[rgb]{ .976,  .447,  .427}57.19} &       & \multicolumn{1}{c}{\cellcolor[rgb]{ .988,  .651,  .467}45.72} & \multicolumn{1}{c}{\cellcolor[rgb]{ .988,  .647,  .467}45.86} & \multicolumn{1}{c}{\cellcolor[rgb]{ .988,  .659,  .467}45.36} &       & \multicolumn{1}{c}{\cellcolor[rgb]{ .98,  .51,  .439}53.65} & \multicolumn{1}{c}{\cellcolor[rgb]{ .98,  .502,  .439}54.01} & \multicolumn{1}{c}{\cellcolor[rgb]{ .98,  .502,  .439}54.06} \\
          & Stochastic gradient & \multicolumn{1}{c}{\cellcolor[rgb]{ .996,  .843,  .506}35.02} &       & \multicolumn{1}{c}{\cellcolor[rgb]{ .976,  .42,  .424}58.53} & \multicolumn{1}{c}{\cellcolor[rgb]{ .973,  .412,  .42}58.97} & \multicolumn{1}{c}{\cellcolor[rgb]{ .976,  .435,  .424}57.86} &       & \multicolumn{1}{c}{\cellcolor[rgb]{ .976,  .459,  .431}56.49} & \multicolumn{1}{c}{\cellcolor[rgb]{ .976,  .467,  .431}56.02} & \multicolumn{1}{c}{\cellcolor[rgb]{ .976,  .463,  .431}56.28} \\
    \midrule
    \multicolumn{11}{l}{\textbf{Inclusion error}} \\
          & Elastic net & \multicolumn{1}{c}{\cellcolor[rgb]{ .482,  .773,  .486}28.49} &       & \multicolumn{1}{c}{\cellcolor[rgb]{ .816,  .867,  .506}32.39} & \multicolumn{1}{c}{\cellcolor[rgb]{ .827,  .871,  .506}32.57} & \multicolumn{1}{c}{\cellcolor[rgb]{ .804,  .863,  .506}32.26} &       & \multicolumn{1}{c}{\cellcolor[rgb]{ .82,  .867,  .506}32.43} & \multicolumn{1}{c}{\cellcolor[rgb]{ .831,  .871,  .506}32.59} & \multicolumn{1}{c}{\cellcolor[rgb]{ .808,  .867,  .506}32.34} \\
          & Gradient boosting & \multicolumn{1}{c}{\cellcolor[rgb]{ .737,  .843,  .502}31.47} &       & \multicolumn{1}{c}{\cellcolor[rgb]{ .831,  .871,  .506}32.59} & \multicolumn{1}{c}{\cellcolor[rgb]{ .847,  .875,  .506}32.77} & \multicolumn{1}{c}{\cellcolor[rgb]{ .816,  .867,  .506}32.41} &       & \multicolumn{1}{c}{\cellcolor[rgb]{ 1,  .922,  .518}34.56} & \multicolumn{1}{c}{\cellcolor[rgb]{ 1,  .918,  .518}34.75} & \multicolumn{1}{c}{\cellcolor[rgb]{ .996,  .918,  .514}34.54} \\
          & Linear regression & \multicolumn{1}{c}{\cellcolor[rgb]{ .459,  .765,  .486}28.20} &       & \multicolumn{1}{c}{\cellcolor[rgb]{ .796,  .863,  .506}32.18} & \multicolumn{1}{c}{\cellcolor[rgb]{ .812,  .867,  .506}32.35} & \multicolumn{1}{c}{\cellcolor[rgb]{ .792,  .859,  .502}32.13} &       & \multicolumn{1}{c}{\cellcolor[rgb]{ .8,  .863,  .506}32.21} & \multicolumn{1}{c}{\cellcolor[rgb]{ .812,  .867,  .506}32.38} & \multicolumn{1}{c}{\cellcolor[rgb]{ .796,  .863,  .506}32.17} \\
          & Logistic classification & \multicolumn{1}{c}{\cellcolor[rgb]{ .388,  .745,  .482}27.36} &       & \multicolumn{1}{c}{\cellcolor[rgb]{ .988,  .667,  .471}46.28} & \multicolumn{1}{c}{\cellcolor[rgb]{ .988,  .663,  .471}46.40} & \multicolumn{1}{c}{\cellcolor[rgb]{ .988,  .682,  .475}45.56} &       & \multicolumn{1}{c}{\cellcolor[rgb]{ 1,  .863,  .506}37.35} & \multicolumn{1}{c}{\cellcolor[rgb]{ 1,  .875,  .51}36.72} & \multicolumn{1}{c}{\cellcolor[rgb]{ 1,  .914,  .518}35.05} \\
          & Naive Bayes & \multicolumn{1}{c}{\cellcolor[rgb]{ .992,  .725,  .482}43.54} &       & \multicolumn{1}{c}{\cellcolor[rgb]{ .992,  .733,  .482}43.28} & \multicolumn{1}{c}{\cellcolor[rgb]{ .992,  .725,  .482}43.60} & \multicolumn{1}{c}{\cellcolor[rgb]{ .992,  .725,  .482}43.54} &       & \multicolumn{1}{c}{\cellcolor[rgb]{ .992,  .718,  .478}43.90} & \multicolumn{1}{c}{\cellcolor[rgb]{ .992,  .718,  .478}43.89} & \multicolumn{1}{c}{\cellcolor[rgb]{ .992,  .718,  .478}43.89} \\
          & Neural network & \multicolumn{1}{c}{\cellcolor[rgb]{ .514,  .78,  .486}28.85} &       & \multicolumn{1}{c}{\cellcolor[rgb]{ .976,  .427,  .424}57.17} & \multicolumn{1}{c}{\cellcolor[rgb]{ .976,  .424,  .424}57.41} & \multicolumn{1}{c}{\cellcolor[rgb]{ .973,  .412,  .42}57.80} &       & \multicolumn{1}{c}{\cellcolor[rgb]{ .988,  .671,  .471}46.06} & \multicolumn{1}{c}{\cellcolor[rgb]{ .988,  .706,  .478}44.53} & \multicolumn{1}{c}{\cellcolor[rgb]{ .992,  .769,  .49}41.66} \\
          & Random forest & \multicolumn{1}{c}{\cellcolor[rgb]{ .91,  .894,  .51}33.52} &       & \multicolumn{1}{c}{\cellcolor[rgb]{ .522,  .784,  .49}28.97} & \multicolumn{1}{c}{\cellcolor[rgb]{ .58,  .8,  .49}29.63} & \multicolumn{1}{c}{\cellcolor[rgb]{ .557,  .792,  .49}29.36} &       & \multicolumn{1}{c}{\cellcolor[rgb]{ .804,  .863,  .506}32.29} & \multicolumn{1}{c}{\cellcolor[rgb]{ .859,  .878,  .506}32.89} & \multicolumn{1}{c}{\cellcolor[rgb]{ .796,  .863,  .506}32.20} \\
          & Stochastic gradient & \multicolumn{1}{c}{\cellcolor[rgb]{ .996,  .78,  .49}41.10} &       & \multicolumn{1}{c}{\cellcolor[rgb]{ .98,  .529,  .443}52.53} & \multicolumn{1}{c}{\cellcolor[rgb]{ .98,  .514,  .439}53.18} & \multicolumn{1}{c}{\cellcolor[rgb]{ .98,  .533,  .443}52.39} &       & \multicolumn{1}{c}{\cellcolor[rgb]{ .988,  .675,  .471}45.92} & \multicolumn{1}{c}{\cellcolor[rgb]{ .992,  .71,  .478}44.31} & \multicolumn{1}{c}{\cellcolor[rgb]{ .992,  .718,  .478}43.87} \\
    \midrule
    \multicolumn{11}{l}{\textbf{Minimum values}} \\
          & \multicolumn{1}{l}{Exclusion error (EE)} & \multicolumn{1}{c}{28.20} &       & \multicolumn{1}{c}{20.44} & \multicolumn{1}{c}{20.40} & \multicolumn{1}{c}{20.14} &       & \multicolumn{1}{c}{22.90} & \multicolumn{1}{c}{23.17} & \multicolumn{1}{c}{23.05} \\
          & \multicolumn{1}{l}{Inclusion error (IE)} & \multicolumn{1}{c}{27.36} &       & \multicolumn{1}{c}{28.97} & \multicolumn{1}{c}{29.63} & \multicolumn{1}{c}{29.36} &       & \multicolumn{1}{c}{32.21} & \multicolumn{1}{c}{32.38} & \multicolumn{1}{c}{32.17} \\
    \midrule
    \multicolumn{11}{l}{(*) Results of machine learning (ML) models without spatial component. Poverty threshold: 40\%} \\
    \end{tabular}%
  \label{tab:smlres}%
\end{table}%

\begin{table}[htbp]
  \centering
  \captionsetup{justification=centering,margin=1cm}
  \caption{Machine learning results with and without spatial analysis, and with and without principal component analysis: exclusion errors and inclusion errors in the test sample. Data from the year 2016 to 2021}
    \begin{tabular}{lrrrrrrrrrr}
    \toprule
    \multicolumn{2}{r}{\multirow{3}[6]{*}{}} & \multicolumn{1}{r}{\multirow{3}[6]{*}{\textbf{\makecell{Benchmark \\ ML models*}}}} &       & \multicolumn{7}{c}{\textbf{Spatial machine learning models (updated data 2016-2021)}} \\
\cmidrule{5-11}    \multicolumn{2}{r}{} &       &       & \multicolumn{3}{c}{\textbf{Without principal component analysis}} &       & \multicolumn{3}{c}{\textbf{With principal component analysis}} \\
\cmidrule{5-7}\cmidrule{9-11}    \multicolumn{2}{r}{} &       &       & \multicolumn{1}{c}{\textbf{4 clusters}} & \multicolumn{1}{c}{\textbf{6 clusters}} & \multicolumn{1}{c}{\textbf{12 clusters}} &       & \multicolumn{1}{c}{\textbf{4 clusters}} & \multicolumn{1}{c}{\textbf{6 clusters}} & \multicolumn{1}{c}{\textbf{12 clusters}} \\
    \midrule
    \multicolumn{11}{l}{\textbf{Exclusion error}} \\
          & \multicolumn{1}{l}{Elastic net} & \multicolumn{1}{c}{\cellcolor[rgb]{ .906,  .894,  .51}29.78} &       & \multicolumn{1}{c}{\cellcolor[rgb]{ .886,  .886,  .51}29.55} & \multicolumn{1}{c}{\cellcolor[rgb]{ .878,  .886,  .51}29.47} & \multicolumn{1}{c}{\cellcolor[rgb]{ .878,  .886,  .51}29.46} &       & \multicolumn{1}{c}{\cellcolor[rgb]{ .886,  .886,  .51}29.56} & \multicolumn{1}{c}{\cellcolor[rgb]{ .882,  .886,  .51}29.50} & \multicolumn{1}{c}{\cellcolor[rgb]{ .878,  .886,  .51}29.46} \\
          & \multicolumn{1}{l}{Gradient boosting} & \multicolumn{1}{c}{\cellcolor[rgb]{ 1,  .918,  .518}31.12} &       & \multicolumn{1}{c}{\cellcolor[rgb]{ .776,  .855,  .502}28.38} & \multicolumn{1}{c}{\cellcolor[rgb]{ .773,  .855,  .502}28.33} & \multicolumn{1}{c}{\cellcolor[rgb]{ .788,  .859,  .502}28.49} &       & \multicolumn{1}{c}{\cellcolor[rgb]{ 1,  .922,  .518}30.81} & \multicolumn{1}{c}{\cellcolor[rgb]{ .988,  .918,  .514}30.65} & \multicolumn{1}{c}{\cellcolor[rgb]{ .992,  .918,  .514}30.69} \\
          & \multicolumn{1}{l}{Linear regression} & \multicolumn{1}{c}{\cellcolor[rgb]{ .867,  .882,  .51}29.32} &       & \multicolumn{1}{c}{\cellcolor[rgb]{ .843,  .875,  .506}29.07} & \multicolumn{1}{c}{\cellcolor[rgb]{ .835,  .875,  .506}29.00} & \multicolumn{1}{c}{\cellcolor[rgb]{ .831,  .871,  .506}28.95} &       & \multicolumn{1}{c}{\cellcolor[rgb]{ .847,  .875,  .506}29.14} & \multicolumn{1}{c}{\cellcolor[rgb]{ .839,  .875,  .506}29.06} & \multicolumn{1}{c}{\cellcolor[rgb]{ .835,  .875,  .506}29.00} \\
          & \multicolumn{1}{l}{Logistic classification} & \multicolumn{1}{c}{\cellcolor[rgb]{ .996,  .843,  .506}35.48} &       & \multicolumn{1}{c}{\cellcolor[rgb]{ .729,  .843,  .502}27.85} & \multicolumn{1}{c}{\cellcolor[rgb]{ .733,  .843,  .502}27.91} & \multicolumn{1}{c}{\cellcolor[rgb]{ .929,  .898,  .51}29.99} &       & \multicolumn{1}{c}{\cellcolor[rgb]{ 1,  .906,  .518}31.75} & \multicolumn{1}{c}{\cellcolor[rgb]{ 1,  .906,  .518}31.90} & \multicolumn{1}{c}{\cellcolor[rgb]{ 1,  .878,  .51}33.32} \\
          & \multicolumn{1}{l}{Naive Bayes} & \multicolumn{1}{c}{\cellcolor[rgb]{ .624,  .812,  .494}26.73} &       & \multicolumn{1}{c}{\cellcolor[rgb]{ .388,  .745,  .482}24.19} & \multicolumn{1}{c}{\cellcolor[rgb]{ .408,  .749,  .482}24.41} & \multicolumn{1}{c}{\cellcolor[rgb]{ .396,  .745,  .482}24.28} &       & \multicolumn{1}{c}{\cellcolor[rgb]{ 1,  .902,  .514}32.13} & \multicolumn{1}{c}{\cellcolor[rgb]{ 1,  .898,  .514}32.17} & \multicolumn{1}{c}{\cellcolor[rgb]{ 1,  .886,  .514}32.91} \\
          & \multicolumn{1}{l}{Neural network} & \multicolumn{1}{c}{\cellcolor[rgb]{ 1,  .906,  .518}31.77} &       & \multicolumn{1}{c}{\cellcolor[rgb]{ 1,  .859,  .506}34.54} & \multicolumn{1}{c}{\cellcolor[rgb]{ 1,  .855,  .506}34.93} & \multicolumn{1}{c}{\cellcolor[rgb]{ .996,  .824,  .502}36.65} &       & \multicolumn{1}{c}{\cellcolor[rgb]{ 1,  .91,  .518}31.58} & \multicolumn{1}{c}{\cellcolor[rgb]{ .867,  .882,  .51}29.36} & \multicolumn{1}{c}{\cellcolor[rgb]{ .886,  .886,  .51}29.56} \\
          & \multicolumn{1}{l}{Random forest} & \multicolumn{1}{c}{\cellcolor[rgb]{ .973,  .412,  .42}60.96} &       & \multicolumn{1}{c}{\cellcolor[rgb]{ .984,  .596,  .455}50.18} & \multicolumn{1}{c}{\cellcolor[rgb]{ .984,  .596,  .455}50.12} & \multicolumn{1}{c}{\cellcolor[rgb]{ .984,  .588,  .455}50.55} &       & \multicolumn{1}{c}{\cellcolor[rgb]{ .98,  .514,  .439}55.03} & \multicolumn{1}{c}{\cellcolor[rgb]{ .98,  .514,  .439}55.06} & \multicolumn{1}{c}{\cellcolor[rgb]{ .98,  .502,  .439}55.73} \\
          & \multicolumn{1}{l}{Stochastic gradient} & \multicolumn{1}{c}{\cellcolor[rgb]{ .992,  .753,  .486}40.79} &       & \multicolumn{1}{c}{\cellcolor[rgb]{ .976,  .439,  .427}59.49} & \multicolumn{1}{c}{\cellcolor[rgb]{ .976,  .431,  .424}59.87} & \multicolumn{1}{c}{\cellcolor[rgb]{ .98,  .498,  .439}55.92} &       & \multicolumn{1}{c}{\cellcolor[rgb]{ .98,  .514,  .439}55.14} & \multicolumn{1}{c}{\cellcolor[rgb]{ .98,  .529,  .443}54.06} & \multicolumn{1}{c}{\cellcolor[rgb]{ .976,  .482,  .435}56.93} \\
    \midrule
    \multicolumn{11}{l}{\textbf{Inclusion error}} \\
          & \multicolumn{1}{l}{Elastic net} & \multicolumn{1}{c}{\cellcolor[rgb]{ .537,  .784,  .49}29.38} &       & \multicolumn{1}{c}{\cellcolor[rgb]{ .804,  .863,  .506}32.70} & \multicolumn{1}{c}{\cellcolor[rgb]{ .8,  .863,  .506}32.67} & \multicolumn{1}{c}{\cellcolor[rgb]{ .804,  .863,  .506}32.73} &       & \multicolumn{1}{c}{\cellcolor[rgb]{ .808,  .863,  .506}32.74} & \multicolumn{1}{c}{\cellcolor[rgb]{ .804,  .863,  .506}32.72} & \multicolumn{1}{c}{\cellcolor[rgb]{ .804,  .863,  .506}32.72} \\
          & \multicolumn{1}{l}{Gradient boosting} & \multicolumn{1}{c}{\cellcolor[rgb]{ .643,  .816,  .494}30.72} &       & \multicolumn{1}{c}{\cellcolor[rgb]{ .71,  .835,  .498}31.57} & \multicolumn{1}{c}{\cellcolor[rgb]{ .71,  .835,  .498}31.55} & \multicolumn{1}{c}{\cellcolor[rgb]{ .725,  .839,  .498}31.72} &       & \multicolumn{1}{c}{\cellcolor[rgb]{ .882,  .886,  .51}33.69} & \multicolumn{1}{c}{\cellcolor[rgb]{ .878,  .886,  .51}33.64} & \multicolumn{1}{c}{\cellcolor[rgb]{ .886,  .886,  .51}33.73} \\
          & \multicolumn{1}{l}{Linear regression} & \multicolumn{1}{c}{\cellcolor[rgb]{ .498,  .776,  .486}28.92} &       & \multicolumn{1}{c}{\cellcolor[rgb]{ .769,  .855,  .502}32.30} & \multicolumn{1}{c}{\cellcolor[rgb]{ .765,  .851,  .502}32.24} & \multicolumn{1}{c}{\cellcolor[rgb]{ .769,  .855,  .502}32.26} &       & \multicolumn{1}{c}{\cellcolor[rgb]{ .773,  .855,  .502}32.35} & \multicolumn{1}{c}{\cellcolor[rgb]{ .769,  .855,  .502}32.29} & \multicolumn{1}{c}{\cellcolor[rgb]{ .773,  .855,  .502}32.32} \\
          & \multicolumn{1}{l}{Logistic classification} & \multicolumn{1}{c}{\cellcolor[rgb]{ .388,  .745,  .482}27.53} &       & \multicolumn{1}{c}{\cellcolor[rgb]{ .988,  .671,  .471}46.30} & \multicolumn{1}{c}{\cellcolor[rgb]{ .988,  .686,  .475}45.61} & \multicolumn{1}{c}{\cellcolor[rgb]{ .988,  .69,  .475}45.45} &       & \multicolumn{1}{c}{\cellcolor[rgb]{ .996,  .82,  .498}39.82} & \multicolumn{1}{c}{\cellcolor[rgb]{ .996,  .839,  .502}38.91} & \multicolumn{1}{c}{\cellcolor[rgb]{ 1,  .859,  .506}37.94} \\
          & \multicolumn{1}{l}{Naive Bayes} & \multicolumn{1}{c}{\cellcolor[rgb]{ .984,  .624,  .463}48.44} &       & \multicolumn{1}{c}{\cellcolor[rgb]{ .984,  .627,  .463}48.38} & \multicolumn{1}{c}{\cellcolor[rgb]{ .984,  .627,  .463}48.34} & \multicolumn{1}{c}{\cellcolor[rgb]{ .984,  .616,  .459}48.85} &       & \multicolumn{1}{c}{\cellcolor[rgb]{ .992,  .722,  .482}44.14} & \multicolumn{1}{c}{\cellcolor[rgb]{ .992,  .725,  .482}43.98} & \multicolumn{1}{c}{\cellcolor[rgb]{ .992,  .71,  .478}44.55} \\
          & \multicolumn{1}{l}{Neural network} & \multicolumn{1}{c}{\cellcolor[rgb]{ .486,  .773,  .486}28.77} &       & \multicolumn{1}{c}{\cellcolor[rgb]{ .976,  .427,  .424}57.16} & \multicolumn{1}{c}{\cellcolor[rgb]{ .976,  .427,  .424}57.12} & \multicolumn{1}{c}{\cellcolor[rgb]{ .973,  .412,  .42}57.81} &       & \multicolumn{1}{c}{\cellcolor[rgb]{ .98,  .533,  .443}52.49} & \multicolumn{1}{c}{\cellcolor[rgb]{ .984,  .588,  .455}50.12} & \multicolumn{1}{c}{\cellcolor[rgb]{ .988,  .639,  .467}47.73} \\
          & \multicolumn{1}{l}{Random forest} & \multicolumn{1}{c}{\cellcolor[rgb]{ 1,  .878,  .51}37.07} &       & \multicolumn{1}{c}{\cellcolor[rgb]{ .718,  .839,  .498}31.63} & \multicolumn{1}{c}{\cellcolor[rgb]{ .686,  .827,  .498}31.23} & \multicolumn{1}{c}{\cellcolor[rgb]{ .765,  .851,  .502}32.21} &       & \multicolumn{1}{c}{\cellcolor[rgb]{ .992,  .918,  .514}35.06} & \multicolumn{1}{c}{\cellcolor[rgb]{ .973,  .914,  .514}34.80} & \multicolumn{1}{c}{\cellcolor[rgb]{ 1,  .922,  .518}35.19} \\
          & \multicolumn{1}{l}{Stochastic gradient} & \multicolumn{1}{c}{\cellcolor[rgb]{ .992,  .776,  .49}41.70} &       & \multicolumn{1}{c}{\cellcolor[rgb]{ .98,  .545,  .447}51.99} & \multicolumn{1}{c}{\cellcolor[rgb]{ .98,  .557,  .451}51.46} & \multicolumn{1}{c}{\cellcolor[rgb]{ .98,  .49,  .435}54.36} &       & \multicolumn{1}{c}{\cellcolor[rgb]{ .984,  .616,  .459}48.90} & \multicolumn{1}{c}{\cellcolor[rgb]{ .988,  .667,  .471}46.51} & \multicolumn{1}{c}{\cellcolor[rgb]{ .992,  .722,  .482}44.05} \\
    \midrule
    \multicolumn{11}{l}{\textbf{Minimum values}} \\
          & Exclusion error (EE) & \multicolumn{1}{c}{26.73} &       & \multicolumn{1}{c}{24.19} & \multicolumn{1}{c}{24.41} & \multicolumn{1}{c}{24.28} &       & \multicolumn{1}{c}{29.14} & \multicolumn{1}{c}{29.06} & \multicolumn{1}{c}{29.00} \\
          & Inclusion error (IE) & \multicolumn{1}{c}{27.53} &       & \multicolumn{1}{c}{31.57} & \multicolumn{1}{c}{31.23} & \multicolumn{1}{c}{31.72} &       & \multicolumn{1}{c}{32.35} & \multicolumn{1}{c}{32.29} & \multicolumn{1}{c}{32.32} \\
    \midrule
    \multicolumn{11}{l}{(*) Machine learning without spatial component} \\
    \end{tabular}%
  \label{tab:smlres_21}%
\end{table}%

\end{landscape}

In the case of average inclusion errors (IE) in the data of 2016 to 2020, similar results are obtained with and without the implementation of spatial contiguity clustering. The lowest average IE (equal to 27\%) in the machine learning implementation without spatial component is obtained with a logistic model, while the lowest IE (29\%) is obtained with random forests in the spatial machine learning implementation. The results for IE suggest that there is room for improving PMT and inclusion errors by developing further the utilised machine learning algorithms or expanding the definition of poverty to incorporate dimensions beyond income like the Multi-dimensional Poverty Index does.

\begin{figure}[ht]
\centering
\captionsetup{justification=centering,margin=1cm}
\caption{Spatial machine learning results at province level: exclusion errors, 4 clusters}
\hspace*{-1in}
\includegraphics[width=5.5in]{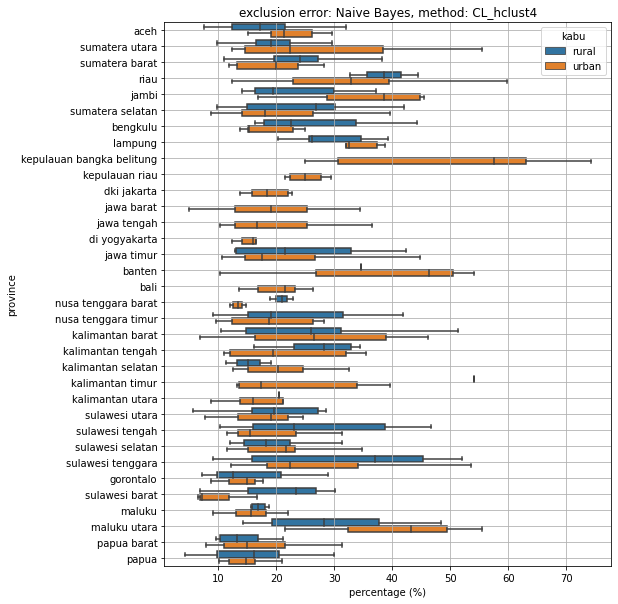}
\label{fig:prov_eepc_hclus4}
\end{figure}

\begin{figure}[ht]
\centering
\captionsetup{justification=centering,margin=1cm}
\caption{Spatial machine learning results at province level: exclusion errors, 6 clusters}
\hspace*{-1in}
\includegraphics[width=5.5in]{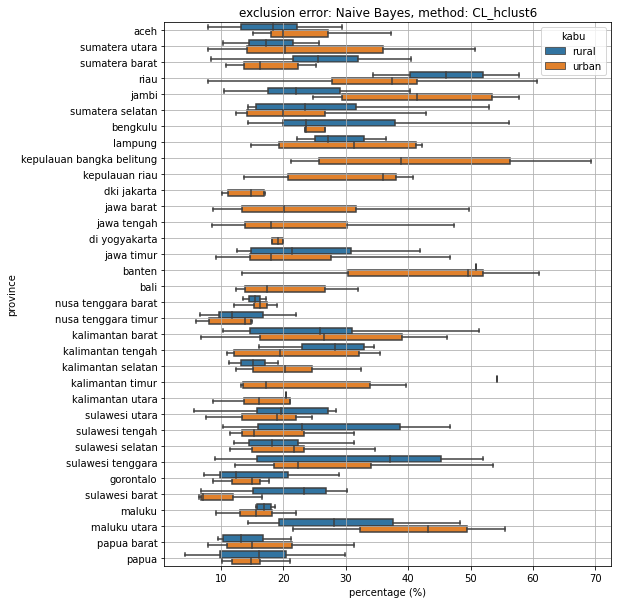}
\label{fig:prov_eepc_hclus6}
\end{figure}

\begin{figure}[ht]
\centering
\captionsetup{justification=centering,margin=1cm}
\caption{Spatial machine learning results at province level: exclusion errors, 12 clusters}
\hspace*{-1in}
\includegraphics[width=5.5in]{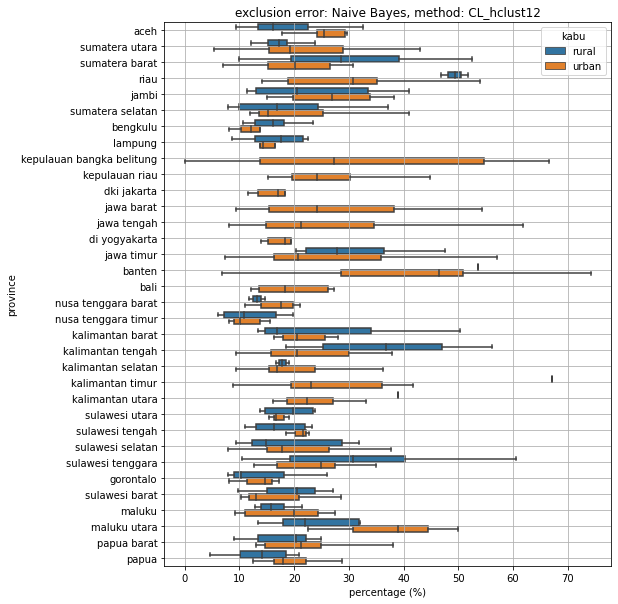}
\label{fig:prov_eepc_hclus12}
\end{figure}

\begin{figure}[ht]
\centering
\captionsetup{justification=centering,margin=1cm}
\caption{Spatial machine learning results at province level: inclusion errors, 4 clusters}
\hspace*{-1in}
\includegraphics[width=5.5in]{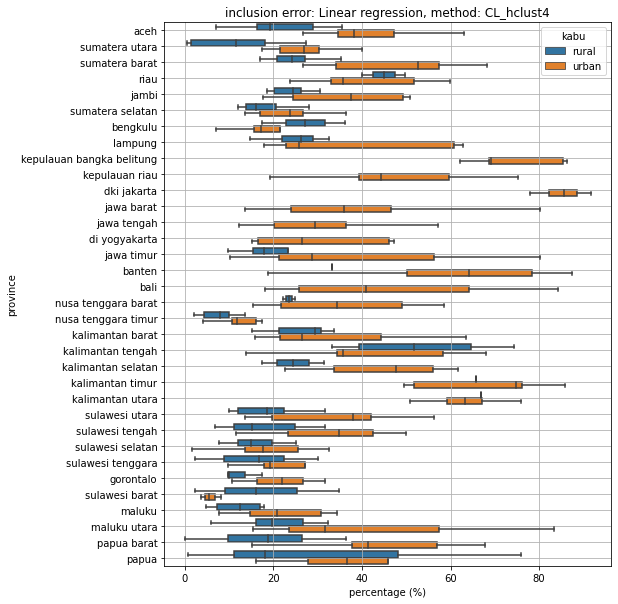}
\label{fig:prov_iepc_hclus4}
\end{figure}

\begin{figure}[ht]
\centering
\captionsetup{justification=centering,margin=1cm}
\caption{Spatial machine learning results at province level: inclusion errors, 6 clusters}
\hspace*{-1in}
\includegraphics[width=5.5in]{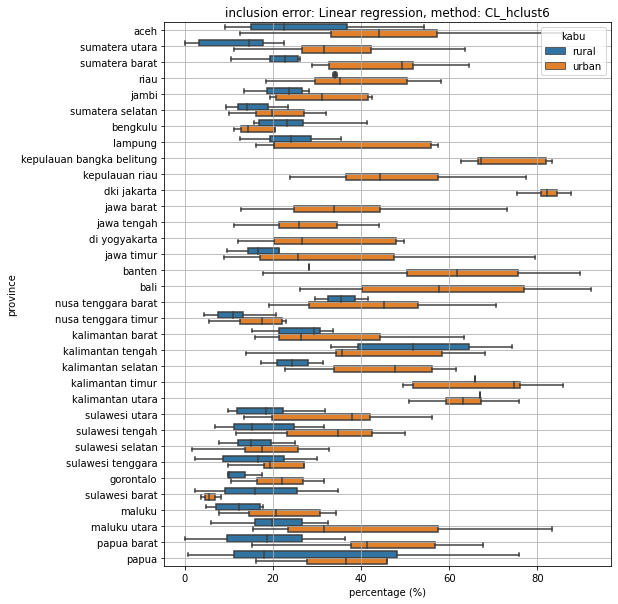}
\label{fig:prov_iepc_hclus6}
\end{figure}

\begin{figure}[ht]
\centering
\captionsetup{justification=centering,margin=1cm}
\caption{Spatial machine learning results at province level: inclusion errors, 12 clusters}
\hspace*{-1in}
\includegraphics[width=5.5in]{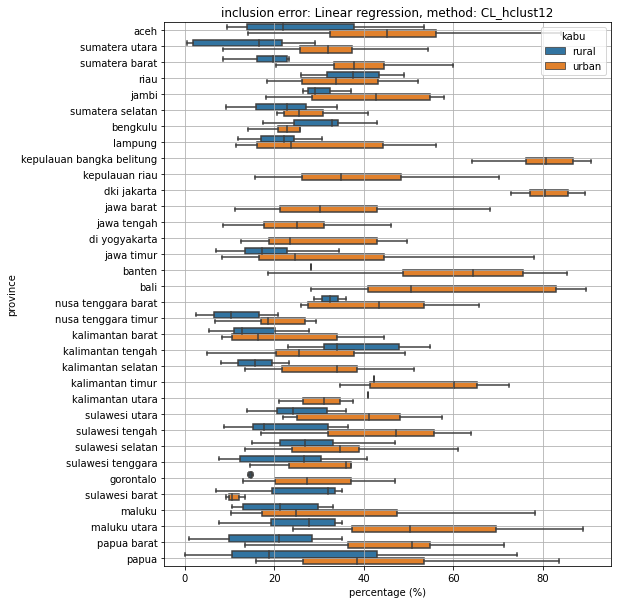}
\label{fig:prov_iepc_hclus12}
\end{figure}

In general, exclusion errors are lower in urban kabupaten compared to rural kabupaten (Figures \ref{fig:prov_eepc_hclus4}, \ref{fig:prov_eepc_hclus6} and \ref{fig:prov_eepc_hclus12}). Urban kabupaten in Lampung, Dki Jakarta or Di Yogyakarta have exclusion errors below 20\% (between 10\% and 20\%), when using 12 spatial clusters (Figure \ref{fig:prov_eepc_hclus12}). Exclusion errors between 10\% and 20\% are also observed in the rural kabupaten for Nusa Tenggara Barat and Nusa Tenggara Timur (Figure \ref{fig:prov_eepc_hclus12}). In the case of inclusion errors (Figures \ref{fig:prov_iepc_hclus4}, \ref{fig:prov_iepc_hclus6} and \ref{fig:prov_iepc_hclus12}), the high average inclusion errors in the overall models are caused by large inclusion errors in provinces such as Kepulauan Riau and Kepulauan Bangka Belitung, while the kabupaten of the province of Sulawesi Barat have inclusion errors below 20\% when using 4, 6 and 12 spatial clusters. The kabupaten of the provinces of Nusa Tenggara Timur and Bengkulu have inclusion errors below 20\% when using 4 and 6 spatial clusters (Appendix 1).

\begin{figure}[ht]
\centering
\captionsetup{justification=centering,margin=1cm}
\caption{Machine learning results without spatial component (benchmark models): inclusion errors}
\includegraphics[width=5in]{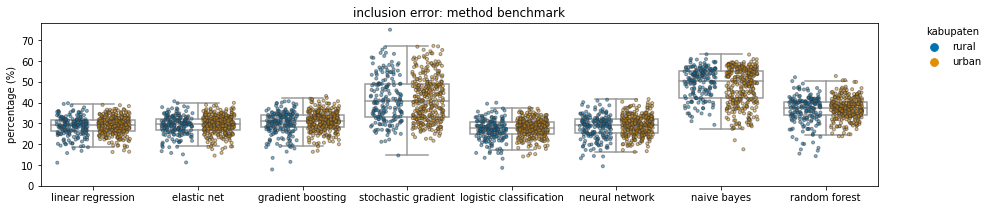}
\label{fig:ie_ben}
\end{figure}


\section{Conclusion}\label{sec:conc}

Improving targeting accuracy in social safety nets is crucial for optimizing resource allocation. Machine learning (ML) techniques have shown superior precision compared to traditional Proxy Means Testing (PMT) methods based on linear regression models. Hybrid, context-specific approaches that integrate ML with geographical information leverage the strengths of each methodology, especially in contexts where evidence of spatial correlation and clustering in household poverty exists. Recently, \citet{corral2025poverty} found that machine learning produces more biased poverty estimates than traditional methods, particularly in the poorest geographic areas. Our study indicates that the poor performance of traditional machine learning models can be linked to a lack of explicit inclusion of spatial contiguity matrices that capture clusters of poverty. We show that spatial machine learning models outperform traditional algorithms for poverty mapping.

In this study, we identified spatial patterns in income levels across Indonesia at both provincial (propinsi) and district (kabupaten) levels. These patterns were used to calculate multiple spatial clusters, which were then incorporated into ML algorithms to enhance the targeting of poor households and reduce exclusion errors in PMT. Our results showed that spatial machine learning models reduced exclusion errors (EE) to 20\% for data from 2016 to 2020 and to 24\% for data from 2016 to 2021, particularly in urban areas. The smaller reduction for 2021 is attributable to the significant number of missing values in this dataset, which were imputed using averages from prior years. These imputed values failed to accurately capture socio-economic changes in income distribution caused by the COVID-19 pandemic. Nonetheless, the observed reductions in EE for both datasets suggest that clustering household data effectively decreases the likelihood of excluding poor households that should qualify for social protection programs in Indonesia.

The models with the lowest exclusion and inclusion errors did not incorporate principal components analysis (PCA). This outcome suggests that feature engineering may outperform statistical dimensionality reduction techniques in reducing targeting errors. Future iterations of poverty-targeting ML algorithms could benefit from a theoretically grounded feature engineering approach. For instance, constructing a multidimensional poverty index (MPI) based on the counting method proposed by Alkire and Foster \citep{alkire2011counting} could improve predictive accuracy. In Indonesia’s context, such an index should incorporate spatial contiguity analysis to further refine poverty targeting estimates.

Future research could explore how incorporating temporal changes in income distributions, particularly during periods of socio-economic disruption that influence the accuracy of poverty targeting models. Investigating the application of spatio-temporal patterns in machine learning could capture more nuanced temporal and spatial dependencies in household income data. 

Improved methods for imputing missing values that account for temporal and spatial variations would ensure more robust predictions in future iterations. Constructing a hybrid index that combines PMT results with other predictive indicators to capture a household's potential vulnerability to shocks such as wih climate change would provide a more comprehensive framework for poverty targeting in the context of a changing environment. By addressing these areas, future research may contribute to the effectiveness and equity of social safety net programs.

\bibliography{refs}

\newpage
\section*{Appendix 1: results at province level (data 2016-2020)}
\label{sec:Appendix}
\addcontentsline{toc}{section}{{Appendix 1: spatial machine learning results at province level (2016-2020)}}

\begin{figure}[ht]
\centering
\captionsetup{justification=centering,margin=1cm}
\caption{Spatial machine learning results at province level: exclusion errors, 4 clusters (no PCA)}
\hspace*{-1in}
\includegraphics[width=5.5in]{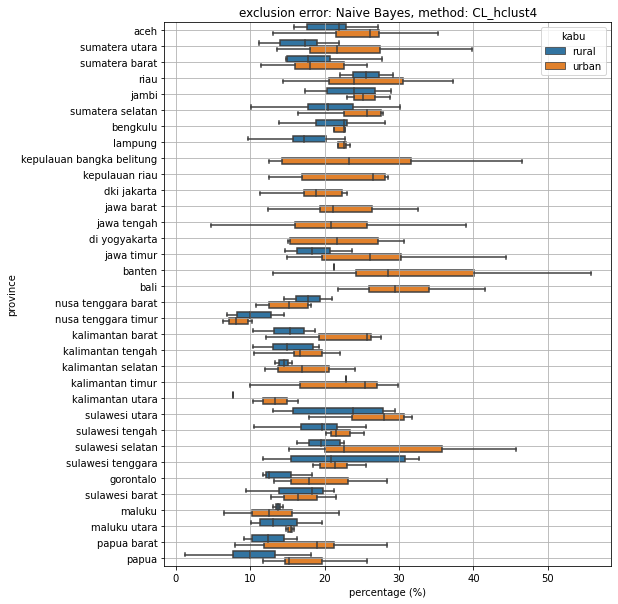}
\label{fig:prov_eenpc_hclus4}
\end{figure}

\begin{figure}[ht]
\centering
\captionsetup{justification=centering,margin=1cm}
\caption{Spatial machine learning results at province level: exclusion errors, 6 clusters (no PCA)}
\hspace*{-1in}
\includegraphics[width=5.5in]{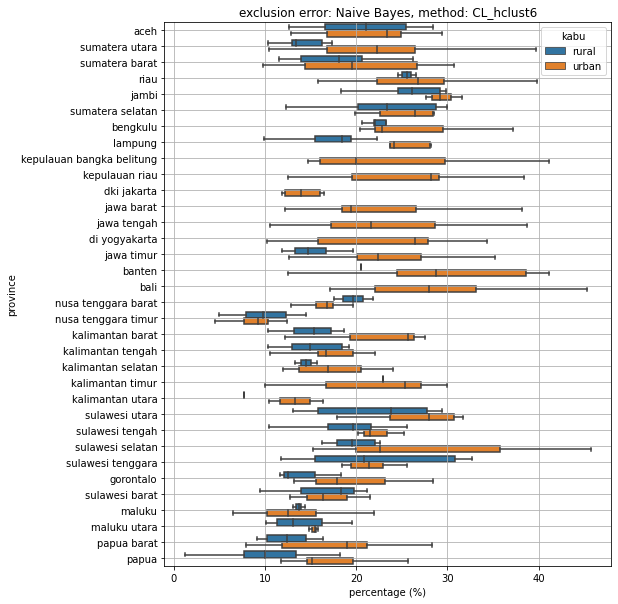}
\label{fig:prov_eenpc_hclus6}
\end{figure}

\begin{figure}[ht]
\centering
\captionsetup{justification=centering,margin=1cm}
\caption{Spatial machine learning results at province level: exclusion errors, 12 clusters (no PCA)}
\hspace*{-1in}
\includegraphics[width=5.5in]{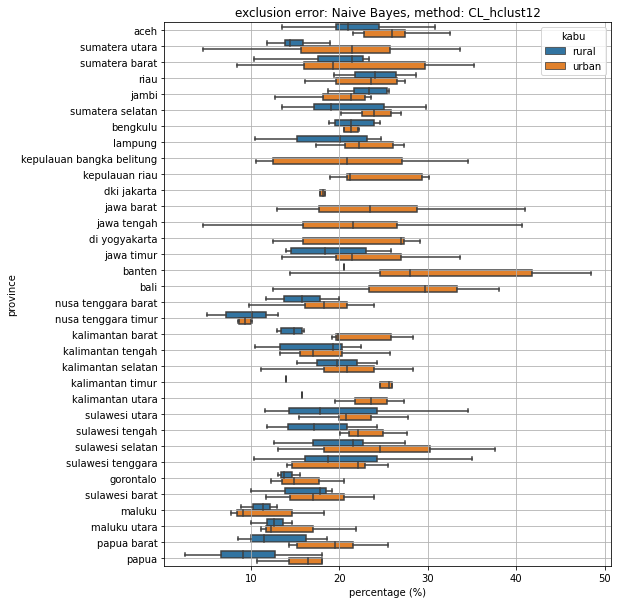}
\label{fig:prov_eenpc_hclus12}
\end{figure}


\begin{figure}[ht]
\centering
\captionsetup{justification=centering,margin=1cm}
\caption{Spatial machine learning results at province level: inclusion errors, 4 clusters (no PCA)}
\hspace*{-1in}
\includegraphics[width=5.5in]{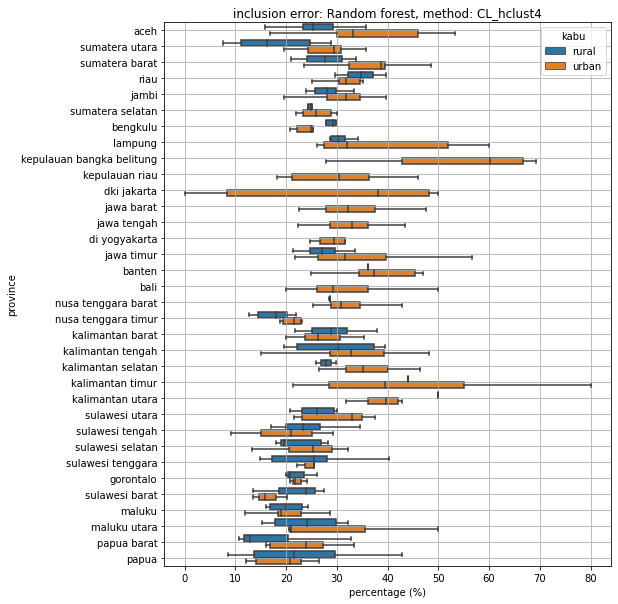}
\label{fig:prov_ienpc_hclus4}
\end{figure}

\begin{figure}[ht]
\centering
\captionsetup{justification=centering,margin=1cm}
\caption{Spatial machine learning results at province level: inclusion errors, 6 clusters (no PCA)}
\hspace*{-1in}
\includegraphics[width=5.5in]{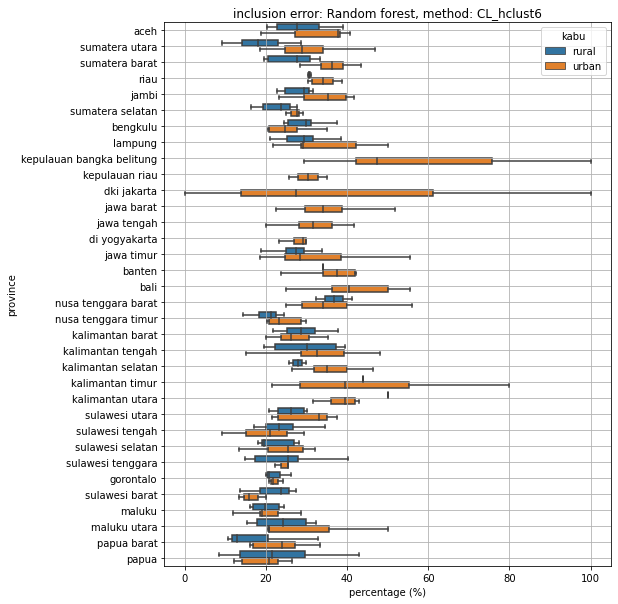}
\label{fig:prov_inepc_hclus6}
\end{figure}

\begin{figure}[ht]
\centering
\captionsetup{justification=centering,margin=1cm}
\caption{Spatial machine learning results at province level: inclusion errors, 12 clusters (no PCA)}
\hspace*{-1in}
\includegraphics[width=5.5in]{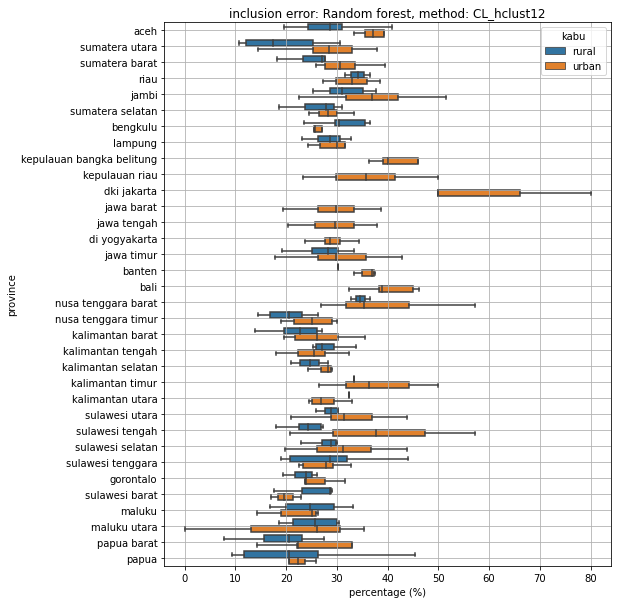}
\label{fig:prov_ienpc_hclus12}
\end{figure}


\section*{Appendix 2: results at province level (data 2016-2021)}
\label{sec:Appendix2}
\addcontentsline{toc}{section}{{Appendix 2: spatial machine learning results at province level (2016-2021)}}

\begin{figure}[ht]
\centering
\captionsetup{justification=centering,margin=1cm}
\caption{Spatial machine learning results at province level: exclusion errors, 4 clusters}
\hspace*{-1in}
\includegraphics[width=5.5in]{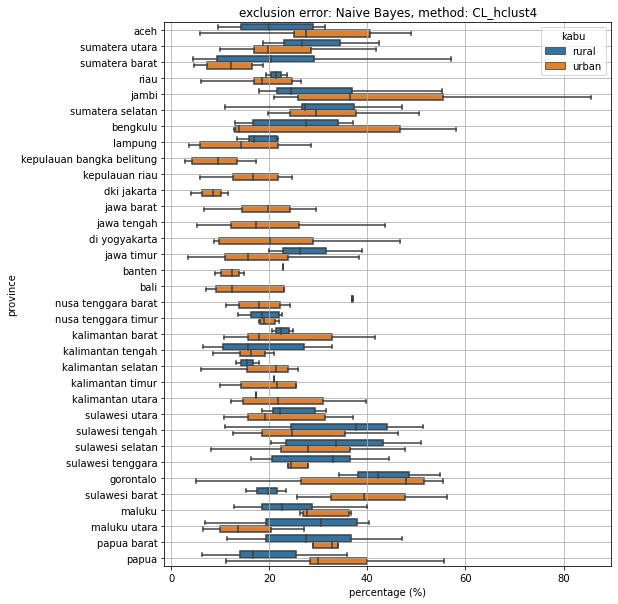}
\end{figure}

\begin{figure}[ht]
\centering
\captionsetup{justification=centering,margin=1cm}
\caption{Spatial machine learning results at province level: exclusion errors, 6 clusters}
\hspace*{-1in}
\includegraphics[width=5.5in]{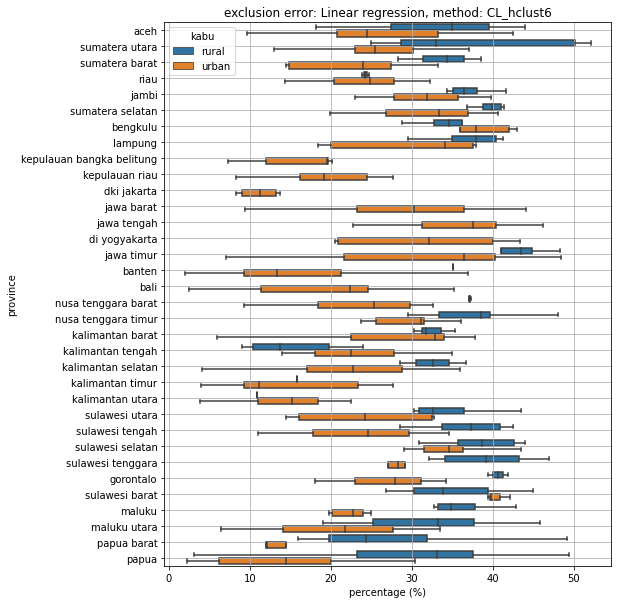}
\end{figure}

\begin{figure}[ht]
\centering
\captionsetup{justification=centering,margin=1cm}
\caption{Spatial machine learning results at province level: exclusion errors, 12 clusters}
\hspace*{-1in}
\includegraphics[width=5.5in]{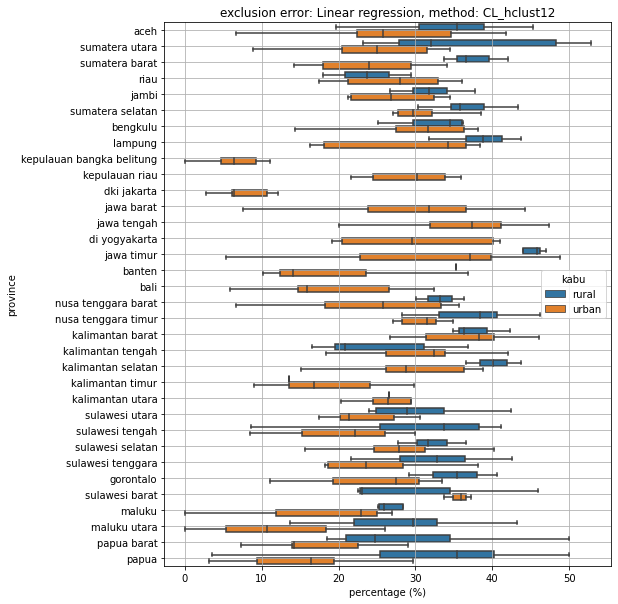}
\end{figure}


\begin{figure}[ht]
\centering
\captionsetup{justification=centering,margin=1cm}
\caption{Spatial machine learning results at province level: inclusion errors, 4 clusters}
\hspace*{-1in}
\includegraphics[width=5.5in]{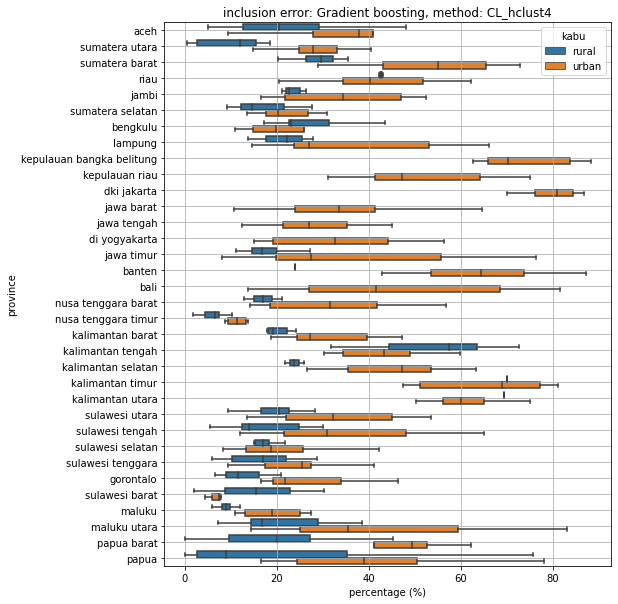}
\end{figure}

\begin{figure}[ht]
\centering
\captionsetup{justification=centering,margin=1cm}
\caption{Spatial machine learning results at province level: inclusion errors, 6 clusters}
\hspace*{-1in}
\includegraphics[width=5.5in]{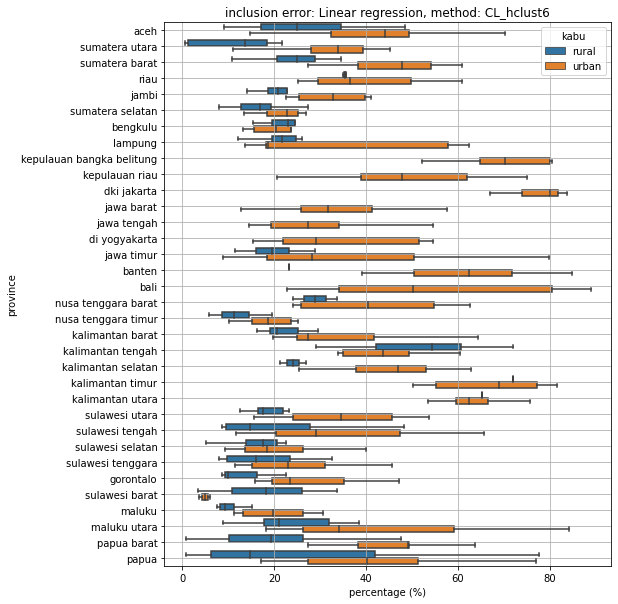}
\end{figure}

\begin{figure}[ht]
\centering
\captionsetup{justification=centering,margin=1cm}
\caption{Spatial machine learning results at province level: inclusion errors, 12 clusters}
\hspace*{-1in}
\includegraphics[width=5.5in]{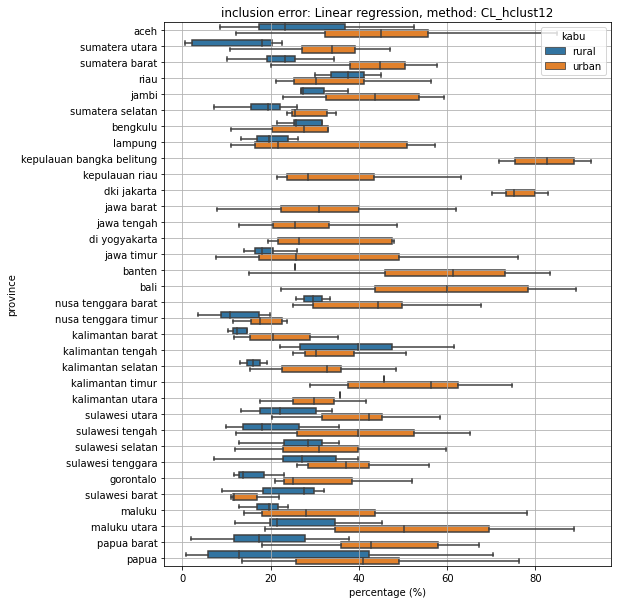}
\end{figure}


\begin{figure}[ht]
\centering
\captionsetup{justification=centering,margin=1cm}
\caption{Spatial machine learning results at province level: exclusion errors, 4 clusters (no PCA)}
\hspace*{-1in}
\includegraphics[width=5.5in]{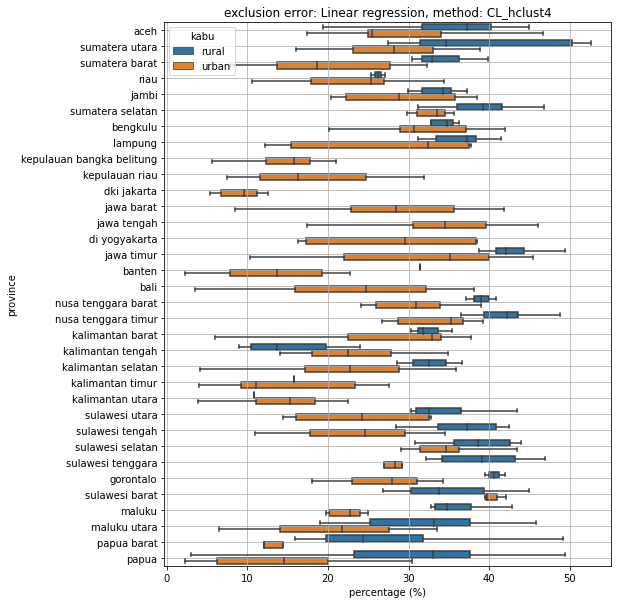}
\end{figure}

\begin{figure}[ht]
\centering
\captionsetup{justification=centering,margin=1cm}
\caption{Spatial machine learning results at province level: exclusion errors, 6 clusters (no PCA)}
\hspace*{-1in}
\includegraphics[width=5.5in]{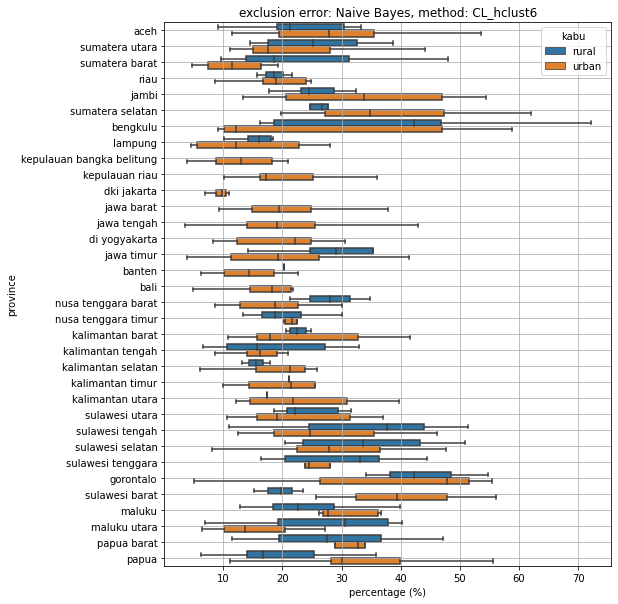}
\end{figure}

\begin{figure}[ht]
\centering
\captionsetup{justification=centering,margin=1cm}
\caption{Spatial machine learning results at province level: exclusion errors, 12 clusters (no PCA)}
\hspace*{-1in}
\includegraphics[width=5.5in]{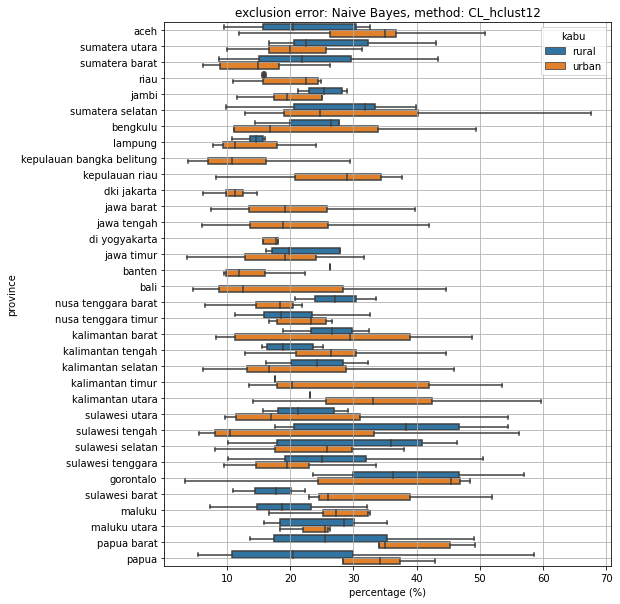}
\end{figure}


\begin{figure}[ht]
\centering
\captionsetup{justification=centering,margin=1cm}
\caption{Spatial machine learning results at province level: inclusion errors, 4 clusters (no PCA)}
\hspace*{-1in}
\includegraphics[width=5.5in]{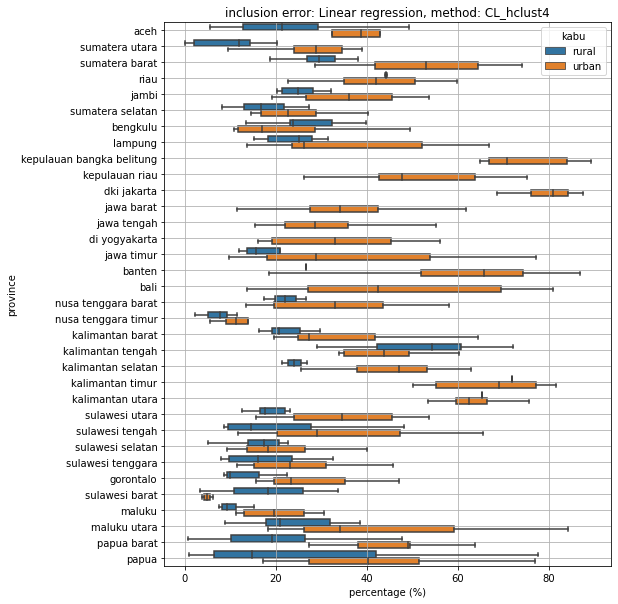}
\end{figure}

\begin{figure}[ht]
\centering
\captionsetup{justification=centering,margin=1cm}
\caption{Spatial machine learning results at province level: inclusion errors, 6 clusters (no PCA)}
\hspace*{-1in}
\includegraphics[width=5.5in]{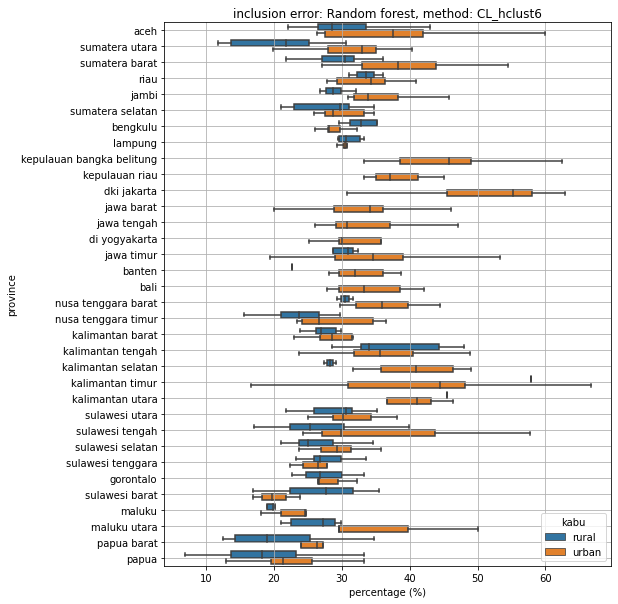}
\end{figure}

\begin{figure}[ht]
\centering
\captionsetup{justification=centering,margin=1cm}
\caption{Spatial machine learning results at province level: inclusion errors, 12 clusters (no PCA)}
\hspace*{-1in}
\includegraphics[width=5.5in]{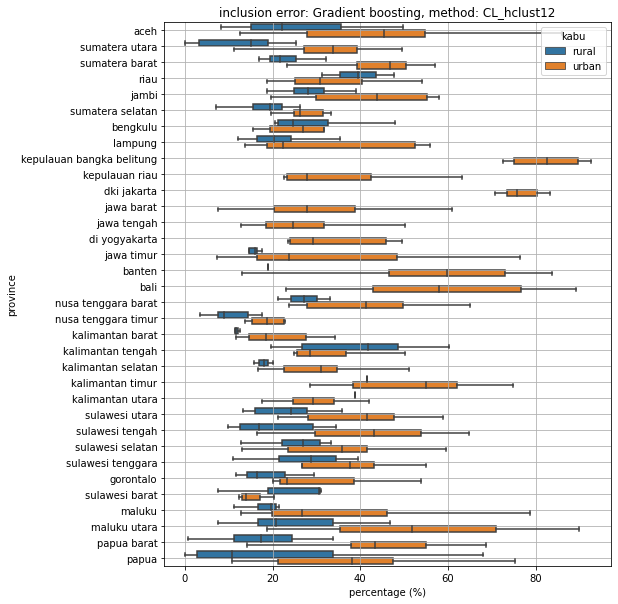}
\end{figure}
\end{document}